  \documentclass{EngC}

  \usepackage[rightcaption,raggedright]{sidecap}% for side captions
  \usepackage{framed}         % for floatingboxes
  \usepackage{soul}           % for letterspacing in theorem-style headings

  \usepackage[agsm]{harvard}

  \usepackage{rotating}
  \usepackage{floatpag}
  \rotfloatpagestyle{empty}

 \usepackage{amsmath}% if you are using this package,
 \usepackage{amssymb}
  \usepackage{amsthm}
  \usepackage{graphicx}
  
  \usepackage{hyperref}
    \usepackage{bm}
  \usepackage{multind}
  \makeindex{authors}
  \makeindex{subject}

  \theoremstyle{plain}% default

  \theoremstyle{definition}

  \theoremstyle{remark}

  \usepackage{array}
  \newcolumntype{L}{>{\centering\arraybackslash}m{2.7cm}}
  \usepackage{multirow}

\usepackage[framemethod=TikZ]{mdframed}
\usepackage{amsthm}
\newcounter{archbox}[section] 
\newcommand{\thetheo}{\arabic{section}.\arabic{archbox}}
\newenvironment{archbox}[2][]{%
\refstepcounter{archbox}%
\ifstrempty{#1}%
{\mdfsetup{%
frametitle={%
\tikz[baseline=(current bounding box.east),outer sep=0pt]
\node[anchor=east,rectangle,fill=blue!20]
{\strut Architecture~\thetheo};}}
}%
{\mdfsetup{%
frametitle={%
\tikz[baseline=(current bounding box.east),outer sep=0pt]
\node[anchor=east,rectangle,fill=gray!20]
{\strut #1};}}%
}%
\mdfsetup{innertopmargin=0pt,linecolor=gray!20,%
linewidth=2pt,topline=true,%
frametitleaboveskip=\dimexpr-\ht\strutbox\relax
}
\begin{mdframed}[]\relax%
\label{#2}}{\end{mdframed}}

\begin{document}

% chap1.tex
% 2011/02/03, v1.10

%%% Jong's macros

\newcommand{\Ab}{{\boldsymbol A}}
\newcommand{\Bb}{{\boldsymbol B}}
\newcommand{\Cb}{{\boldsymbol{C}}}
\newcommand{\Db}{{\boldsymbol D}}
\newcommand{\Eb}{{\boldsymbol E}}
\newcommand{\Fb}{{\boldsymbol F}}
\newcommand{\Gb}{{\boldsymbol G}}
\newcommand{\Hb}{{\boldsymbol H}}
\newcommand{\Ib}{{\boldsymbol I}}
\newcommand{\Jb}{{\boldsymbol J}}
\newcommand{\Kb}{{\boldsymbol K}}
\newcommand{\Lb}{{\boldsymbol L}}
\newcommand{\Mb}{{\boldsymbol M}}
\newcommand{\Nb}{{\boldsymbol N}}
\newcommand{\Ob}{{\boldsymbol O}}
\newcommand{\Pb}{{\boldsymbol P}}
\newcommand{\Qb}{{\boldsymbol Q}}
\newcommand{\Rb}{{\boldsymbol R}}
\newcommand{\Sb}{{\boldsymbol S}}
\newcommand{\Tb}{{\boldsymbol T}}
\newcommand{\Ub}{{\boldsymbol U}}
\newcommand{\Vb}{{\boldsymbol V}}
\newcommand{\Wb}{{\boldsymbol W}}
\newcommand{\Xb}{{\boldsymbol X}}
\newcommand{\Yb}{{\boldsymbol Y}}
\newcommand{\Zb}{{\boldsymbol Z}}

\newcommand{\Sigmab}{{\boldsymbol {\Sigma}}}

\newcommand{\xb}{{\boldsymbol x}}
\newcommand{\yb}{{\boldsymbol y}}
\newcommand{\mb}{{\boldsymbol m}}
\newcommand{\zb}{{\boldsymbol z}}
\newcommand{\Rc}{\mathcal{R}}

% sentence
\newcommand{\Hs}{\mathscr{H}}
\newcommand{\Xs}{\mathscr{X}}
\newcommand{\Hc}{\mathcal{H}}
\newcommand{\Mc}{\mathcal{M}}
\newcommand{\Ac}{\mathcal{A}}
\newcommand{\Bc}{\mathcal{B}}
\newcommand{\Cc}{\mathcal{C}}
\newcommand{\Lc}{\mathcal{L}}
\newcommand{\Nc}{\mathcal{N}}
\newcommand{\Ic}{\mathcal{I}}

\newcommand{\Tc}{\mathcal{T}}
\newcommand{\Xc}{\mathcal{X}}
\newcommand{\Yc}{\mathcal{Y}}
\newcommand{\Sc}{\mathcal{S}}
\newcommand{\Vc}{\mathcal{V}}
\newcommand{\Wc}{\mathcal{W}}
\newcommand{\Zc}{\mathcal{Z}}

\newcommand{\1}{\boldsymbol{1}}

\newcommand{\Rd}{{\mathbb R}}
\newcommand{\add}[1] {\textcolor{k}{#1}} % for changes by JCY

\chapter{Deep learning for ultrasound beamforming}
\label{DLFUB}
\vspace{-2cm}
\textbf{Ruud JG van Sloun, Jong Chul Ye and Yonina C Eldar}

\section{Introduction and relevance}
\noindent Diagnostic imaging plays a critical role in healthcare, serving as a fundamental asset for timely diagnosis, disease staging and management as well as for treatment choice, planning, guidance, and follow-up. Among the diagnostic imaging options, ultrasound imaging \cite{szabo2004diagnostic} is uniquely positioned, being a highly cost-effective modality that offers the clinician an unmatched and invaluable level of interaction, enabled by its real-time nature. Its portability and cost-effectiveness permits point-of-care imaging at the bedside, in emergency settings, rural clinics, and developing countries. Ultrasonography is increasingly used across many medical specialties, spanning from obstetrics, cardiology and oncology to acute and intensive care, with a market share that is globally growing.

On the technological side, ultrasound probes are becoming increasingly compact and portable, with the market demand for low-cost `pocket-sized' devices (i.e. ``the stethoscope model'') expanding \cite{baran2009design}. Transducers are miniaturized, allowing e.g. in-body imaging for interventional applications. At the same time, there is a strong trend towards 3D imaging \cite{provost20143d} and the use of high-frame-rate imaging schemes \cite{tanter2014ultrafast}; both accompanied by dramatically increasing data rates that pose a heavy burden on the probe-system communication and subsequent image reconstruction algorithms. Systems today offer a wealth of advanced applications and methods, including shear wave elasticity imaging \cite{bercoff2004supersonic}, ultra-sensitive Doppler \cite{demene2015spatiotemporal}, and ultrasound localization microscopy for super-resolution microvascular imaging \cite{errico2015ultrafast}, \cite{christensen2020super}.

With the demand for high-quality image reconstruction and signal extraction from less (e.g unfocused or parallel) transmissions that facilitate fast imaging, and a push towards compact probes, modern ultrasound imaging leans heavily on innovations in powerful digital receive channel processing. Beamforming, the process of mapping received ultrasound echoes to the spatial image domain, naturally lies at the heart of the ultrasound image formation chain. In this chapter, we discuss why and when deep learning methods can play a compelling role in the digital beamforming pipeline, and then show how these data-driven systems can be leveraged for improved ultrasound image reconstruction \cite{vanSloun2019DL_in_US}.

This chapter is organized as follows. Sec.~\ref{sec:imageformation} briefly introduces various scanning modes in ultrasound. Then, in Sec.~\ref{sec:digitalbeamforming}, we describe methods and rational for digital receive beamforming. In Sec.~\ref{sec:opportunities} we elaborate on the opportunities of deep learning for ultrasound beamforming, and in Sec.~\ref{sec:architectures} we review various deep network architectures. We then turn to typical approaches for training in Sec.~\ref{sec:training}. Finally, in Sec.~\ref{sec:new} we discuss several future directions.

\section{Ultrasound scanning in a nutshell}
\label{sec:imageformation}
\noindent Ultrasound imaging is based on the pulse-echo principle. First, a radiofrequency (RF) pressure wave is transmitted into the medium of interest through a multi-element ultrasound transducer. These transducers are typically based on piezoelectric (preferably single-crystal) mechanisms or CMUT technology. After insonification, the acoustic wave backscatters due to inhomogeneities in the medium properties, such as density and speed of sound. The resulting reflections are recorded by the same transducer array and used to generate a so-called `brightness-mode' (B-mode) image through a signal processing step termed beamforming. We will elaborate on this step in Sec.~\ref{sec:digitalbeamforming}. The achievable resolution, contrast, and overall fidelity of B-mode imaging depends on the array aperture and geometry, element sensitivity and bandwidth. Transducer geometries include linear, curved or phased arrays. The latter is mainly used for extended field-of-views from limited acoustic windows (e.g. imaging of the heart between the ribs), enabling the use of angular beam steering due to a smaller pitch, namely, distance between elements. The elements effectively sample the aperture: using a pitch of half the wavelength (i.e. spatial Nyquist rate sampling) avoids grating lobes (spatial aliasing) in the array response. 2D ultrasound imaging is based on 1D arrays, while 3D imaging makes use of 2D matrix designs.

\begin{figure}
    \centering
    \includegraphics[scale=0.6]{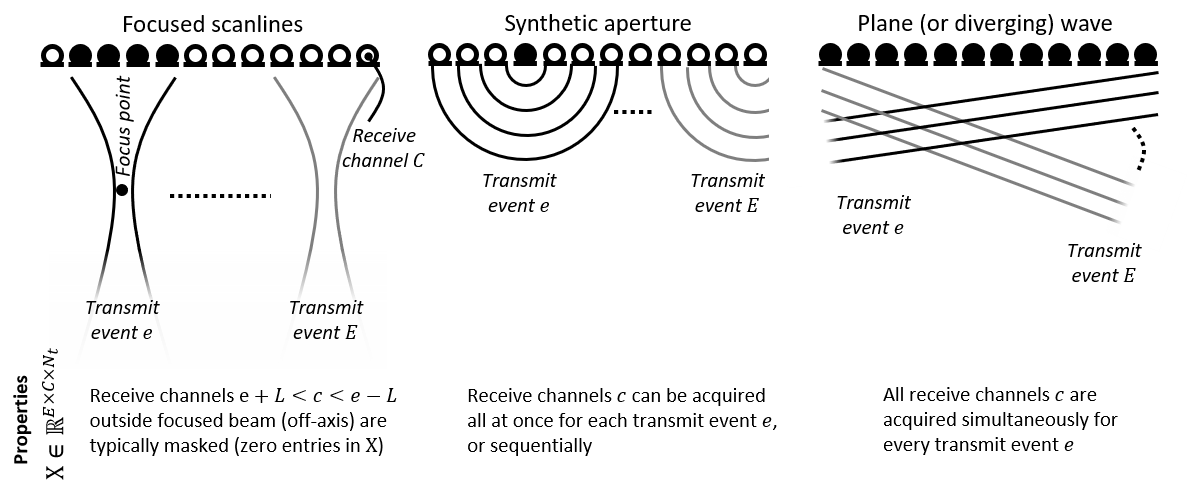}
    \caption{An illustration of three transmit types. Transmit events are denoted by $e$, receive channels as $c$ and, for focused scanlines, $2L+1$ is the size of the active aperture in terms of elements.}
    \label{fig:transmit}
\end{figure}

Given the transducer's physical constraints, getting the most out of the system requires careful optimization across its entire imaging chain. At the front-end, this starts with the design of appropriate transmit schemes for wave field generation. At this stage, crucial trade-offs are made, in which the frame rate, imaging depth, and attainable axial and lateral resolution are weighted carefully against each other: improved resolution can be achieved through the use of higher pulse modulation frequencies and bandwidths; yet, these shorter wavelengths suffer from increased absorption and thus lead to reduced penetration depth. Likewise, high frame rate can be reached by exploiting parallel transmission schemes based on e.g. planar or diverging waves. However, use of such unfocused transmissions comes at the cost of loss in lateral resolution compared to line-based scanning with tightly focused beams. As such, optimal transmit schemes depend on the application. We will briefly elaborate on three common transmit schemes for ultrasound B-mode imaging, an illustration of which is given in Fig.~\ref{fig:transmit}.

\subsection{Focused transmits / line scanning}
In line scanning, a series of $E$ transmit events is used to, with each transmit $e$, produce a single depth-wise line in the image by focusing the transmitted acoustic energy along that line. Such focused transmits are typically achieved using a subaperture of transducer elements $c\in\lbrace e-L,..,e+L \rbrace$, excited with time-delayed RF pulses. By choosing these transmit delays per channel appropriately, the beam can be focused towards a given depth and (for phased arrays) angle. Focused line scanning is the most common transmit design in commercial ultrasound systems, enjoying improved lateral resolution and image contrast compared to unfocused transmits. Line-by-line acquisition is however time consuming (every lateral line requires a distinct transmit event), upper bounding the frame rate of this transmit mode by the number of lines, imaging depth, and speed of sound. This constraint can be relaxed via multi-line parallel transmit approaches, at the expense of reduced image quality.

\subsection{Synthetic aperture}
Synthetic aperture transmit schemes allow for synthetic dynamic transmit focusing by acquiring echoes with the full array following near-spherical wave excitation using individual transducer elements $c$. By performing $E$ such transmits (typically $E=C$, the number of array elements), the image reconstruction algorithm (i.e. the beamformer) has full access to all transmit-receive pairs, enabling retrospective focusing in both transmit and receive. Sequential transmissions with all elements is however time consuming, and acoustic energy delivered into the medium is limited (preventing e.g. harmonic imaging applications).
Synthetic aperture imaging also finds application in phased array intravascular ultrasound (IVUS) imaging, where one can only transmit and receive using one element/channel at a time due to catheter-based constraints.

\subsection{Plane wave / ultrafast}
Today, an increasing amount of ultrasound applications rely on high frame-rate (dubbed \textit{ultrafast}) parallel imaging based on plane waves or diverging waves. Among these are e.g. ultrasound localization microscopy, highly-sensitive Doppler, shear wave elastography and (blood) speckle tracking. Where the former two mostly exploit the incredible vastness of data to obtain accurate signal statistics, the later two leverage high-speed imaging to track ultrasound-induced shear waves or tissue motion to estimate elasticity, strain or flow parameters. In plane wave imaging, planar waves are transmitted to insonify the full region of interest in a single transmit. Typically, several plane waves with different angles are compounded to improve image quality. For small-footprint phased arrays, diverging waves are used. These diverging waves are based on a set of virtual focus points that are placed behind the array, acting as virtual spherical sources. In that context, one can also interpret diverging wave imaging as a synthetic aperture technique.

With the expanding use of ultrafast transmit sequences in modern ultrasound imaging, a strong burden is placed on the subsequent receive channel processing. High data-rates not only raise substantial hardware complications related to data storage and data transfer, in addition, the corresponding unfocused transmissions require advanced receive beamforming to reach satisfactory image quality.

\section{Digital ultrasound beamforming}
\label{sec:digitalbeamforming}
We now describe how the received and digitized channel data is used to reconstruct an image in the digital domain, via a digital signal processing algorithm called beamforming.

\subsection{Digital beamforming model and framework}
\begin{figure}
    \centering
    \includegraphics[scale=0.8]{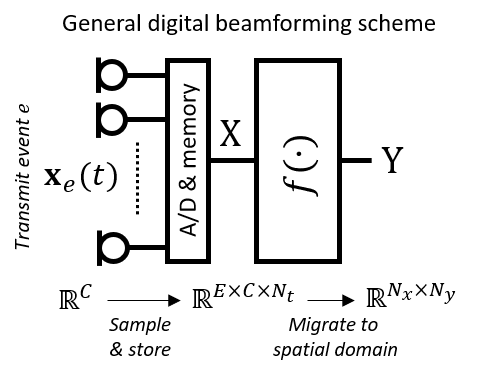}
    \caption{General beamforming scheme.}
    \label{fig:BF}
\end{figure}

%Something about 2D/3D
Consider an array of $C$ channels, and $E$ transmit events, resulting in $E\times C$ measured and digitized RF signal vectors containing $N_t$ samples. Denote $\textrm{X} \in \mathbb{R}^{E\times C\times N_t}$ as the resulting received RF data cube. Individual transmit events $e$ can be tilted planar waves, diverging waves, focused scanlines through transmit beamforming, or any other desired pressure distribution in transmission. The goal of beamforming is to map this time-array-domain `channel' data cube to the spatial domain, through a processor $f(\cdot)$:
\begin{equation}
    \label{eqn:beamforming}
    \mathrm{Y} = f(\mathrm{X}),
\end{equation}
where $\mathrm{Y}\in \mathbb{R}^{N_x\times N_y}$ denotes the beamformed spatial data, with $N_x$ and $N_y$ being the number of pixels in the axial and lateral direction, respectively. In principle, all beamforming architectures in ultrasound can be formulated according to \eqref{eqn:beamforming} and the illustration in Fig. \ref{fig:BF}. The different approaches vary in their parameterization of $f(\cdot)$. Most are composed of a geometrical time-to-space migration of individual channels, and a subsequent combiner/processor. We will now go over some of the most common parameterizations for ultrasound beamforming. In Sec.~\ref{sec:architectures}, we will see that these conventional signal processing methods can also directly inspire parameterizations comprising deep neural networks.

\subsection{Delay-and-sum}
\begin{figure}
    \centering
    \includegraphics[scale=0.8]{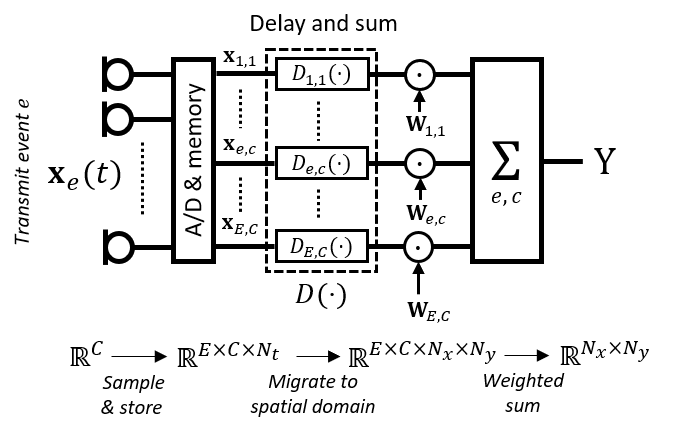}
    \caption{Delay-and-sum beamforming.}
    \label{fig:DAS}
\end{figure}

The industry standard beamforing algorithm is delay-and-sum beamforming (DAS). DAS is commonplace due to its low complexity, allowing for real-time image reconstruction at the expense of non-optimal image quality. Its processing can in general be written as (see Fig.~\ref{fig:DAS} for an illustration):
\begin{equation}
    \label{eqn:beamforming}
    \mathrm{Y} = \sum_{E,L} \mathrm{W}\odot D(\mathrm{X}),
\end{equation}
where $\mathrm{W}\in \mathbb{R}^{E\times C \times N_x \times N_y}$ is an apodization weight tensor and $\odot$ denotes the element-wise product. Note that apodization weights can be complex valued when $D(\mathrm{X})$ is IQ demodulated, allowing for phase shifting by $\mathrm{W}$. In the remainder of this chapter, we will (without loss of generality) only consider real weights applied to RF data. Here, $D(\cdot)$ is a focussing function that migrates the RF time signal to space for each transmit event and channel, mapping $\mathrm{X}$ from $\mathbb{R}^{E\times C \times N_t}$ to $\mathbb{R}^{E\times C \times N_x \times N_y}$. This mapping is obtained by applying geometry-based time delays to the RF signals with the aim of time-aligning the received echoes from the set of focal points.

DAS is typically employed in a so-called dynamic receive beamforming mode, in which the focal points change as a function of scan depth. In a specific variant of dynamic receive beamforming, \textit{pixel-based beamforming}, each pixel is a focus point. Note that unlike its name suggests, dynamic does not mean that the beamformer is dynamically updating $D(\cdot)$ and $\mathrm{W}$ on the fly. Its name stems from the varying time-delays across fast time\footnote{In ultrasound imaging a distinction is made between slow-time and fast-time: slow-time refers to a sequence of snapshots (i.e., across multiple transmit/receive events), at the pulse repetition rate, whereas fast-time refers to the time axis of the received RF signal for a given transmit event.} (and therewith depth) to dynamically move the focal point deeper.

As said, channel delays are used to time-align the received echoes from a given position, and are determined by ray-based wave propagation, dictated by the array geometry, transmit design (plane wave, diverging wave, focused, SA), position of interest, and an estimate of the speed of sound. For each focal point $\{r_x,r_y\}$, channel $c$, and transmit event $e$, we can write the time-of-flight as:
\begin{equation}
    \tau_{e,c,r_x,r_y} = \tau_{e,c,\mathbf{r}} = \frac{||\mathbf{r}_{e}-\mathbf{r}||_2+||\mathbf{r}_{c}-\mathbf{r}||_2}{v},
\end{equation}
where $\tau_{e,c,\mathbf{r}}$ is the time-of-flight for an imaging point $\mathbf{r}$, $\mathbf{r}_{c}$ is the position vector of the receiving element in the array, and $v$ is the speed of sound in the medium. The vector $\mathbf{r}_{e}$ depends on the transmit sequence: for focused transmits it is the position vector of the (sub)aperture center for transmit $e$; for synthetic aperture it is the position vector of the $e^\text{th}$ transmitting element; for diverging waves it is the position vector of the $e^\text{th}$ virtual source located behind the array; for plane waves this is dictated by the transmit angle.

For any focus point $\{r_x,r_y\}$, the response at channel $c$ for a given transmit event $e$ is thus given by:
\begin{equation}
\mathbf{z}_{e,c}[r_x,r_y] = D_{e,c}(\mathbf{x}_{e,c};\tau_{e,c,r_x,r_y}),
\label{eqn:delay}
\end{equation}
where $\mathbf{x}_{e,c}$ denotes the received signal for the $e^\text{th}$ transmit event and $c^\text{th}$ channel, and $D_{e,c}(\mathbf{x};\tau)$ migrates $\mathbf{x}_{e,c}$ from time to space based on the geometry-derived delay $\tau$. To achieve high-resolution delays in the discrete domain, \eqref{eqn:delay} is typically implemented using interpolation or fractional delays with polyphase filters.  Alternatively, delays can be implemented in the Fourier domain, which, as we shall discuss later, has practical advantages for e.g. compressed sensing applications \cite{chernyakova2014fourier}.

After migrating $\mathrm{X}$ to the spatial domain the apodization tensor $\mathrm{W}$ is applied, and the result is (coherently) summed across the $e$ (transmit event) and $c$ (channel) dimensions. Design of the apodization tensor $\mathrm{W}$ inherently poses a compromise between main lobe width and side lobe intensity, or equivalently, resolution and contrast/clutter. This can be intuitively understood from the far-field Fourier relationship between the beampattern and array aperture: analogous to filtering in frequency, the properties of a beamformer (spatial filter) are dictated by the sampling (aperture pitch), filter length (aperture size), and coefficients (apodization weights). Typical choices include Hamming-style apodizations, suppressing sidelobes at the expense of a wider main lobe. In commercial systems, the full (depth/position-dependent) weight tensor is carefully engineered and fine-tuned based on the transducer design (e.g. pitch, size, center frequency, near/far-field zones) and imaging application (e.g. cardiac, obstetrics, general imaging, or even intravascular).

\subsection{Adaptive beamforming}
Adaptive beamforming aims to overcome the inherent tradeoff between sidelobe levels and resolution of static DAS beamforming by making its apodization weight tensor $\mathrm{W}$ fully data-adaptive, i.e., $\mathrm{W} \triangleq  \mathrm{W}(\mathrm{X})$. Adaptation of $\mathrm{W}$ is based on estimates of the array signal statistics, which are typically calculated instantaneously on a spatiotemporal block of data, or through recursive updates. Note that in general, the performance of these methods is bounded by the bias and variance of these statistics estimators. The latter can be reduced by using a larger block of samples (assuming some degree of spatio-temporal stationarity), which comes at the cost of reduced spatiotemporal adaptivity, or techniques such as sub-array averaging.

We will now briefly review some typical adaptive beamforming structures. In general, we distinguish between beamformers that act on individual channels and those that use the channel statistics to compute a single weighting factor across all channels, so called postfilters.\\
\\
\\
\noindent\textbf{Minimum variance}\\
A popular adaptive beamforming method is the minimum variance distortionless response (MVDR), or Capon, beamformer. The MVDR beamformer acts on individual channels, with the optimal weights $\mathrm{W}\in \mathbb{R}^{E\times C \times N_x \times N_y}$ defined as those that, for each transmit event $e$ and location $[r_x,r_y]$, minimize the total signal variance/power while maintaining distortionless response in the direction/focal-point of interest. This amounts to solving:
\begin{equation}
\label{eqn:beamforming_capon}
\begin{aligned}
\hat{\mathbf{w}}_{mv,e}[r_x,r_y] = \underset{\mathbf{w}}{\arg\min} \;&\mathbf{w}^H \mathbf{R}_{\mathrm{x}_e[r_x,r_y]} \mathbf{w} \\
\text{s.t. } &\mathbf{w}^H \mathbf{1}=1,
\end{aligned}
\end{equation}
where $\hat{\mathbf{w}}_{mv,e}[r_x,r_y]\in\mathbb{R}^{C} $ is the weight vector for a given transmit event and location, and $\mathbf{R}_{\mathrm{x}_e[r_x,r_y]}$ denotes the estimated channel covariance matrix for transmit event $e$ and location $[r_x,r_y]$. Solving \eqref{eqn:beamforming_capon} requires inversion of the covariance matrix, which grows cubically with the number of array channels. This makes MV beamforming computationally much more demanding than DAS, in particular for large arrays. This in practice results in a significantly longer reconstruction time and thereby deprives ultrasound of the interactability that makes it so appealing compared to e.g. MRI and CT. To boost image quality, eigen-space based MV beamforming \cite{asl2010eigenspace} performs an eigendecomposition of the covariance matrix and subsequent signal subspace selection before inversion. This further increases computational complexity. While significant progress has been made to decrease the computational time of MV beamforming algorithms \cite{Kim2014}\cite{Bae2016}, real-time implementation remains a major challenge. In addition, it relies on accurate estimates of the signal statistics, which (as mentioned above) requires some form of spatiotemporal averaging.\\
\\
\noindent\textbf{Coherence factor (CF)}\\
Coherence Factor (CF) weighing \cite{Mallart1994} also applies content-adaptive apodization weights $\mathrm{W}$, however with a specific structure: the weights across different channels are identical/tied. CF weighing thus in practice acts as a post-filter after DAS beamforming. The pixel-wise weighing in this post-filter is based on a ``coherence factor'': the ratio between the coherent and incoherent energy across the array channels. CF weighting however suffers from artifacts when the SNR is low and estimation of coherent energy is challenging \cite{Nilsen2010}. This is particularly problematic for unfocused techniques such as PW and SA imaging. \\
\\
\\
\noindent\textbf{Wiener}\\
The Wiener beamformer produces a minimum mean-squared-error (MMSE) estimate of the signal amplitude $A$ stemming from a particular direction/location of interest:
\begin{equation}
\label{eqn:beamforming_Wiener}
\begin{aligned}
\hat{\mathbf{w}}_e[r_x,r_y] = \underset{\mathbf{w}}{\arg\min}  E(| A - \mathbf{w}^H\mathbf{x}_e[r_x,r_y]|^2).
\end{aligned}
\end{equation}
The solution is \cite{nilsen2010wiener}:
\begin{equation}
\label{eqn:beamforming_Wienersolution}
\begin{aligned}
\hat{\mathbf{w}}_e[r_x,r_y] = \frac{|A|^2}{|A|^2+(\mathbf{w}_{mv,e}[r_x,r_y])^H\mathrm{R}_{n_{e}[r_x,r_y]} \mathbf{w}_{mv,e}[r_x,r_y]} \mathbf{w}_{mv,e}[r_x,r_y],
\end{aligned}
\end{equation}
where $\mathrm{R}_{n_{e}[r_x,r_y]}$ is the noise covariance matrix. Wiener beamforming is thus equivalent to the MVDR beamformer followed by a (CF-like) post-filter that scales the output as a function of the remaining noise power after MVDR beamforming (second term in the denominator). Note that the Wiener beamformer requires estimates of both the signal power and noise covariance matrix. The latter can e.g. be estimated by assuming i.i.d. white noise, i.e., $\mathrm{R}_{n_{e}[r_x,r_y]}=\sigma_n^2 \mathrm{I}$, and calculating $\sigma_n$ from the mean squared difference between the MVDR beamformed output and the channel signals. \\
\\
\noindent\textbf{Iterative Maximum-a-Posteriori (iMAP)}\\

\noindent Chernyakova \textit{et al.} propose an iterative maximum-a-posteriori (iMAP) estimator \cite{Chernyakova2019}, which formalizes post-filter-based methods as a MAP problem by incorporating a statistical prior for the signal of interest. Assuming a zero-mean Gaussian random variable of interest with variance $\sigma_a^2$ that is uncorrelated to the noise (also Gaussian, with variance $\sigma^2_{n}$), the beamformer can be derived as:
\begin{equation}
\begin{aligned}
\hat{\mathbf{w}}_e[r_x,r_y] = \frac{\sigma_a[r_x,r_y]^2}{M\sigma_a[r_x,r_y]^2+M\sigma_{n}[r_x,r_y]^2} \mathbf{1}.
\end{aligned}
\label{eqn:imap}
\end{equation}
The signal and noise variances are estimated in an iterative fashion: first a beamformed output is produced according to \eqref{eqn:imap}, and then the noise variance is estimated based on the mean-squared difference with the individual channels. This process is repeated until a stopping criterion is met. Note that while iMAP performs iterative estimation of the statistics, the beamformer itself shares strong similarity with Wiener postfiltering and CF beamforming.

\begin{figure}
    \centering
    \includegraphics[scale=0.8]{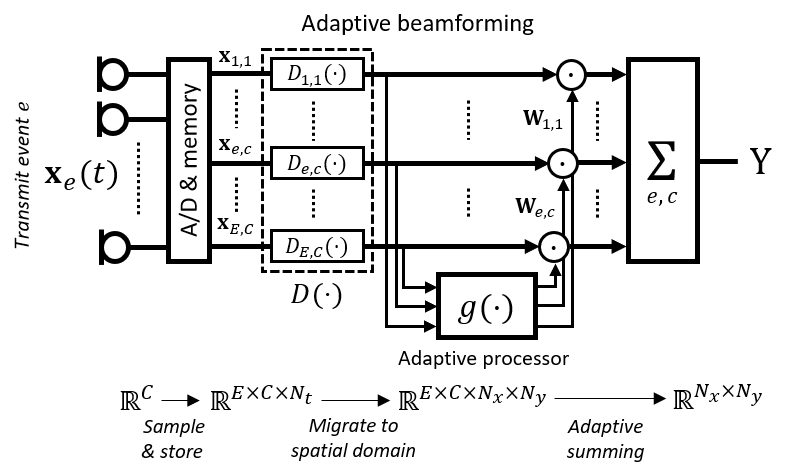}
    \caption{Adaptive beamforming}
    \label{fig:Adaptive}
\end{figure}

%\subsection{Fourier domain beamforming}
%\label{sec:fourierbeamforming}
%\begin{figure}
%    \centering
%    \includegraphics[scale=0.8]{Figures/FourierBeamforming.png}
%    \caption{Fourier domain beamforming}
%    \label{fig:Fourier}
%\end{figure}

\section{Deep learning opportunities}
\label{sec:opportunities}
In this section we will elaborate on some of the fundamental challenges of ultrasound beamforming, and the role that deep learning solutions can play in overcoming these challenges \cite{vanSloun2019DL_in_US}.

\subsection{Opportunity 1: Improving image quality}
As we saw in the previous section, classic adaptive beamformers are derived based on specific modeling assumptions and knowledge of the signal statistics. This limits the performance of model-based adaptive beamformers, which are bounded by:
\begin{enumerate}
    \item Accuracy and precision of the estimated signal statistics using data sampled from a strongly non-stationary ultrasound RF process. In practice, only limited samples are available, making accurate estimation challenging.
    \item Adequacy of the simple linear acquisition model (including homogeneous speed of sound) and assumptions on the statistical structure of desired signals and noise (uncorrelated Gaussian). In practice speed of sound is heterogeneous and noise statistics are highly complex and correlated with signal, e.g. via multiple scattering that leads to reverberation and haze in the image.
\end{enumerate}
In addition, specifically for MVDR beamforming, the complexity of the required matrix inversions hinders real-time implementation. Deep learning can play an important role in addressing these issues, enabling:
\begin{enumerate}
    \item Drastic acceleration of slow model-based approaches such as MVDR, using accelerated neural network implementations as function approximators.
    \item High-performant beamforming outputs without explicit estimation of signal statistics by exploiting useful priors learned from previous examples (training data).
    \item Data-driven nonlinear beamforming architectures that are not bounded by modelling assumptions on linearity and noise statistics. While analytically formalizing the underlying statistical models of ultrasound imaging is highly challenging, and its optimization likely intractable, deep learning circumvents this by learning powerful models directly from data.
\end{enumerate}

\subsection{Opportunity 2: Enabling fast and robust compressed sensing}
Performing beamforming in the digital domain requires sampling the signals received at the transducer elements and transmitting the samples to a back-end processing unit. To achieve sufficient delay resolution for focusing, hundreds of channel signals are typically sampled at 4-10 times their bandwidth, i.e., the sampling rate may severely exceed the Nyquist rate. This problem becomes even more pressing for 3D ultrasound imaging based on matrix transducer technology, where directly streaming channel data from thousands of sensors would lead to data rates of thousands of gigabits per second. Today's technology thus relies on microbeamforming or time-multiplexing to keep data rates manageable. The former compresses data from multiple (adjacent) transducer elements into a single line, thereby virtually reducing the number of receive channels and limiting the attainable resolution and image quality. The latter only communicates a subset of the channel signals to the backend of the system for every transmit event, yielding reduced frame rates.

To overcome these data-rate challenges without compromising image quality and frame rates, a significant research effort has been focused on compressed sensing for ultrasound imaging. Compressed sensing permits low-data-rate sensing (below the Nyquist rate) with strong signal recovery guarantees under specific conditions \cite{eldar2012compressed,eldar2015sampling}. In general, one can perform compressed sensing along three axes in ultrasound: 1) fast-time, 2) slow-time, and 3) channels/array elements. We denote the undersampled measured RF data cube as $X_u\in \mathbb{R}^{E_u \times C_u \times N_u}$, with $E_uC_uN_u < ECN$.  \\
\\
\noindent\textbf{Sub-Nyquist fast-time sampling}\\
To perform sampling rate reduction across fast-time, one can consider the received signals within the framework of finite rate of innovation (FRI) \cite{eldar2015sampling,gedalyahu2011multichannel}. Tur \textit{et al.} \cite{tur2011innovation} modeled the received signal at each element as a finite sum of replicas of the transmitted pulse backscattered from reflectors. The replicas are fully described by their unknown amplitudes and delays, which can be recovered from the signals' Fourier series coefficients. The latter can be computed from low-rate samples of the signal using compressed sensing (CS) techniques \cite{eldar2015sampling,eldar2012compressed}. In \cite{wagner2011xampling,wagner2012compressed}, the authors extended this approach and introduce compressed ultrasound beamforming. It was shown that the beamformed signal follows an FRI model and thus it can be reconstructed from a linear combination of the Fourier coefficients of the received signals. Moreover, these coefficients can be obtained from low-rate samples of the received signals taken according to the Xampling framework \cite{mishali2011xampling2,mishali2011xampling,michaeli2012xampling}. Chernyakova \textit{et al.} showed this Fourier domain relationship between the beam and the received signals holds irrespective of the FRI model. This leads to a general concept of frequency domain beamforming (FDBF) \cite{chernyakova2014fourier} which is equivalent to beamforming in time. FDBF allows to sample the received signals at their effective Nyquist rate without assuming a structured model, thus, it avoids the oversampling dictated by digital implementation of beamforming in time. When assuming that the beam follows a FRI model, the received signals can be sampled at sub-Nyquist rates, leading to up to 28 fold reduction in sampling rate, i.e. $N_t = 28N_u$ \cite{chernyakova2018fourier,burshtein2016sub,lahav2017focus}.\\
\\
\noindent\textbf{Channel and transmit-event compression}\\
Significant research effort has also been invested in exploration of sparse array designs ($C_u<C$) \cite{liu2017maximally,cohen2018beamforming} and efficient sparse sampling of transmit events ($E_u<E$). It has been shown that with proper sparse array selection and a process called convolutional beamforming, the beampattern can be preserved using far fewer elements than the standard uniform linear array \cite{cohen2018sparse}. Typical designs include sparse periodic arrays \cite{austeng2002sparse} or fractal arrays \cite{cohen2020sparse}. In \cite{besson2016compressed_1} a randomly sub-sampled set of receive transducer elements is used, and the work in \cite{lorintiu2015compressed} proposes learned dictionaries for improved CS-based reconstruction from sub-sampled RF lines. In \cite{huijben2020learning}, the authors use deep learning to optimize channel selection and slow-time-sampling for B-mode and downstream color Doppler processing, respectively. Reduction in both time sampling and spatial sampling can be achieved by Compressed Fourier-Domain Convolutional Beamforming leading to reductions in data of two orders of magnitude \cite{mamistvalov2020compressed}.\\
\\

\noindent\textbf{Beamforming and image recovery after compression}\\
After compressive acquisition, dedicated signal recovery algorithms are used to perform image reconstruction from the undersampled dataset. Before deep learning became popular, these algorithms relied on priors/regularizers (e.g. sparsity in some domain) to solve the typically ill-posed optimization problem in a model-based (iterative) fashion. They assume knowledge about the measurement PSF and other system parameters. However, as mentioned before, the performance of model-based algorithms is bounded by the accuracy of the modelling assumptions, including the acquisition model and statistical priors. In addition, iterative solvers are time-consuming, hampering real-time implementation. Today, deep learning is increasingly used to overcome these challenges \cite{perdios2017deep,kulkarni2016reconnet}: 1) Deep learning can be used to learn complex statistical models (explicitly or implicitly) directly from training data, 2) Neural-networks can serve as powerful function approximators that accelerate iterative model-based implementations.

\subsection{Opportunity 3: Beyond MMSE with task-adaptive beamforming}
Classically, beamforming is posed as a signal recovery problem under spatially white Gaussian noise. In that context, optimal beamforming is defined as the beamformer that best estimates the signal of interest in a minimum mean squared error (MMSE) sense. However, beamforming is rarely the last step in the processing chain of ultrasound systems. It is typically followed by demodulation (envelope detection), further image enhancement, spatiotemporal processing (e.g. motion estimation), and image analysis. In that regard it may be more meaningfull to define optimality of the beamformer with respect to its downstream task. We refer this as task-adaptive beamforming. We can define several such tasks in ultrasound imaging. First, we have tasks that focus on further enhancement of the images after beamforming. This includes downstream processing for e.g. de-speckling, de-convolution, or super-resolution, which all have different needs and requirements from the beamformer output. For instance, the performance of deconvolution or super-resolution algorithms is for a large part determined by the invertibility of the point-spread-function model, shaped by the beamformer. Beyond image enhancement, one can think of motion estimation tasks (requiring temporal consistency of the speckle patterns with a clear spatial signature that can be tracked), or even further downstream applications such as segmentation or computer aided diagnosis (CAD).

One may wonder how to optimize beamforming for such tasks in practice. Fortunately, many of these downstream processing tasks are today performed using convolutional neural networks or derivatives therefrom. If the downstream processor is indeed a neural architecture or another algorithm through which one can easily backpropagate gradients, one can directly optimize a neural-network beamformer with respect to its downstream objective/loss through deep learning methods. In this case, \textit{deep} not only refers to the layers in individual networks, but also the stack of beamforming and downstream neural networks. If backpropagation is non-trivial, one could resort to Monte-Carlo gradient estimators based on e.g. the REINFORCE estimator \cite{williams1992simple} or more generally through reinforcement learning algorithms \cite{sutton2018reinforcement}.

\subsection{A brief overview of the state-of-the-art}
The above opportunities have spurred an ever growing collection of papers from the research community. In Table~\ref{table:literature}, we provide an overview of a selection of these in the context of these opportunities, and the challenges they aim to address. Many of these papers simultaneously address more than one challenge/opportunity.

\begin{table}[h!]
\begin{tabular}{LLLL}
& \multicolumn{3}{c}{\large\textbf{Opportunity and focus}} \\ \vspace{1pt}
%\cline{2-4} \\
\textbf{References} & \textbf{Real-time high image quality} \textit{(Main goals)} & \textbf{Compressed sensing} \textit{(Subsampling axis)}  & \textbf{Task-adaptive} \textit{(Considered tasks)}\\
\vspace{1pt} \\
\hline
\vspace{1pt} \\
\cite{luchies2018deep} & Reduce off-axis scattering & - & -\\
\cite{yoon2018efficient} & - & Transmit and channels & - \\
\cite{khan2020adaptive} & - & Channels & -\\
\cite{khan2020switchable} & Boost resolution or suppress speckle & - & Deconvolution and speckle reduction \\
\cite{Luijten2019beamforming} & Boost contrast and resolution & Transmit and Channels & - \\
\cite{nair2020deep} & - & - & Segmentation \\
\cite{wiacek2020coherenet} & Accelerate coherence imaging & - & - \\
\cite{kessler2020deep} and \cite{mamistvalov2021deep} & - & Channels and fast-time Fourier coefficients & - \\
\cite{huijben2020learning} & - & Channels and slow-time & Doppler and B-Mode\\
\cite{hyun2019beamforming} & Suppress speckle & - & Speckle reduction \\
\hline
\end{tabular}
\caption{\label{table:literature}Overview of some of the current literature and their main focus in terms of the opportunities defined in section \ref{sec:opportunities}.}
\end{table}

% related others
% nair2018fully: same as nair but for IUS
% nair2018deep: precursor of nair2020deep
% luchies2019training: training improvements over luchies2018deep
% luchies2020assessing: robustness assessment of luchies 2018deep

\subsection{Public datasets and open source code}
To support the development of deep-learning-based solutions in the context of these opportunities, a challenge on ultrasound beamforming by deep learning (CUBDL) was organized. For public raw ultrasound channel datasets as well as open source code we refer the reader to the challenge website \cite{challenge} and the paper describing the datasets, methods and tools \cite{challengepaper}.

\section{Deep learning architectures for ultrasound beamforming}
\label{sec:architectures}

\subsection{Overview and common architectural choices}
Having set the scope and defined opportunities, we now turn to some of the most common implementations of deep learning in ultrasound beamforming architectures. As in computer vision, most neural architectures for ultrasound beamforming are based on 2D convolutional building blocks. With that, they thus rely on translational equivariance/spatial symmetry of the input data. It is worth noting that this symmetry only holds to some extent for ultrasound imaging, as its point spread function in fact changes as a function of location. When operating on raw channel data before time-space migration (TOF correction), these effects become even more pronounced. Most architectures also restrict the receptive field of the neural network, i.e. beamforming outputs for given spatial location (line/pixel) are computed based on a selected subset of $\mathrm{X}$. This is either implicit, through the depth and size of the selected convolutional kernels, or explicit, by for example only providing a selected number of depth slices \cite{khan2020adaptive} as an input to the network.

In the following we will discuss a number of architectures. We explicitly specify their input, architectural design choices, and output to clarify what part of the beamforming chain they are acting on and how. We point out that the below is not exhaustive, but with examples selected to illustrate and cover the spectrum of approaches and design choices from an educational perspective.

%\subsection{Deep learning architectures for beamforming}
%Define the functions, training  methods etc. mathematically, so we can %refer to them in the following sections.

%\subsubsection{Inputs/receptive field}
%Aperture domain inputs? Pixel-wise? Fourier? etc.
%Compressed inputs

\vspace{1cm}

\subsection{DNN directly on channel data}
In \cite{nair2020deep}, deep learning is used directly on the raw channel data. The deep neural network thus has to learn both the classically geometry-based time-to-space migration (TOF correction) as well as the subsequent beamsumming of channels to yield a beamformed image. In particular the former makes this task particularly challenging - the processor is not a typical image-to-image mapping but rather a time-to-space migration. In addition to yielding a beamformed output, the network in parallel also provides a segmentation mask, which is subsequently used to enhance the final image by masking regions in the image that are classified as anechoic.
\vspace{3cm}
\begin{archbox}[DNN directly on channel data by Nair et al. ]{arch:nair}
\noindent\textbf{Acquisition type}: Single plane wave imaging.

\noindent\textbf{Input}: Complex IQ demodulated data cube $\mathrm{X}_{in}\in \mathbb{C}^{1 \times C \times N_t}$, reformatted to real inputs $\mathrm{X}_{in}\in \mathbb{R}^{C \times N_t \times 2}$ before being fed to the network.

\noindent\textbf{Architecture}:
U-net variant consisting of a single VGG encoder and two parallel decoders with skip connections that map to B-mode and segmentation outputs respectively. The encoder has 10 convolutional layers (kernel size = $3\times 3$) with batch normalization and downsamples the spatial domain via $2\times 2$ max pooling layers while simultaneously increasing the number of feature channels. The two decoders both comprise 9 convolutional layers and perform spatial upsampling to map the feature space back to the desired spatial domain.

\noindent\textbf{Output}: RF B-mode data and pixel-wise class probabilities (segmentation) for the full image $\mathrm{Y}_{bf}\in \mathbb{R}^{N_x \times N_y}$, and $\mathrm{Y}_{seg}\in \mathbb{R}^{N_x \times N_y \times 1}$, respectively.
\end{archbox}

\subsection{DNN for beam-summing}
\noindent\textbf{Hybrid architectures: geometry and learning} \\
While the work by \cite{nair2020deep} replaces the entire beamformer by a deep neural network, most of today's embodiments of ultrasound beamforming with deep learning work on post-delayed channel data. That is, the migration from time-to-space is a deterministic pre-processing step that relies on geometry-based TOF correction. This holds for all of the specific architectures that we will cover in the following. In that sense, they are all hybrid model-based/data-driven beamforming architectures. \\

\begin{figure}
    \centering
    \includegraphics[scale=0.8]{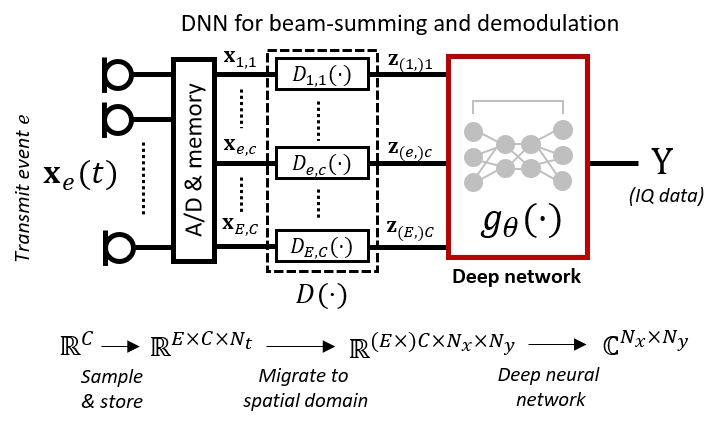}
    \caption{DNN replacing the beam-summing processor.}
    \label{fig:DNN_demod}
\end{figure}

\noindent\textbf{Learning improved beam-summing} \\
We will now discuss designs that replace only the beam-summing stage by a deep network, i.e. after TOF correction, as illustrated in Fig.~\ref{fig:DNN_demod}. In \cite{yoon2018efficient,Khan2019,khan2020adaptive,khan2020switchable,vignon2020resolution}, the authors apply deep convolutional neural networks to perform improved channel aggregation/beam-summing after TOF correction.

\begin{archbox}[DNN beamsumming by Khan et al. ]{arch:khan}
\noindent\textbf{Acquisition type}: Line scanning, so the first dimension (transmit events / lines $E$) has been mapped to the lateral dimension $N_y$ during the time-space migration/TOF correction.

\noindent\textbf{Input}: For each depth/axial location, a TOF-corrected RF data cube $\mathrm{Z}_{in}\in \mathbb{R}^{C \times 3 \times N_y}$. The input data cube comprises a stack of 3 axial slices centered around the axial location of interest.

\noindent\textbf{Architecture}:
37 convolutional layers (kernel size = $3\times 3$), of which all but the last have batch normalization and ReLU activations.

\noindent\textbf{Output}: IQ data for each axial location $\mathbf{y}_{n_x}\in \mathbb{R}^{2 \times 1 \times N_y}$, where the first dimension contains the in-phase (I) and quadrature (Q) components, such that we can also define $\mathbf{y}_{n_x}\in \mathbb{C}^{N_x \times 1}$.

\end{archbox}

\cite{kessler2020deep} use a similar strategy, albeit that pre-processing TOF correction is performed in the Fourier domain. This allows efficient processing when sensing only a small set of the Fourier coefficients of the received channel data. The authors then use a deep convolutional network to perform beamsumming, mapping the TOF-corrected channel data into a single beamformed RF image without aliasing artefacts:
\begin{archbox}[DNN beamsumming by Kessler et al. ]{arch:kessler}
\noindent\textbf{Acquisition type}: Phased array line scanning, so the first dimension (transmit events / angular lines $E$) has been mapped to the lateral dimension $N_y$ during the time-space migration/TOF correction. Reconstruction is in the polar domain, i.e. $N_x$ refers to radial position, and $N_y$ to angular position.

\noindent\textbf{Input}: TOF-corrected RF data cube $\mathrm{Z}_{in}\in \mathbb{R}^{C \times N_x \times 3}$. The input data cube comprises data corresponding to 3 angles centered around the angular position of interest.

\noindent\textbf{Architecture}:
U-net variant with 3 contracting blocks and 3 expanding blocks of convolutional layers with parametric ReLU (PReLU) activations.

\noindent\textbf{Output}: RF data for each angle of interest $\mathbf{y}_{n_y}\in \mathbb{R}^{N_x \times 1}$.

\end{archbox}

%\noindent\textbf{Coherence imaging} \\
%\cite{wiacek2020coherenet}

\subsection{DNN as an adaptive processor}
\begin{figure}
    \centering
    \includegraphics[scale=0.8]{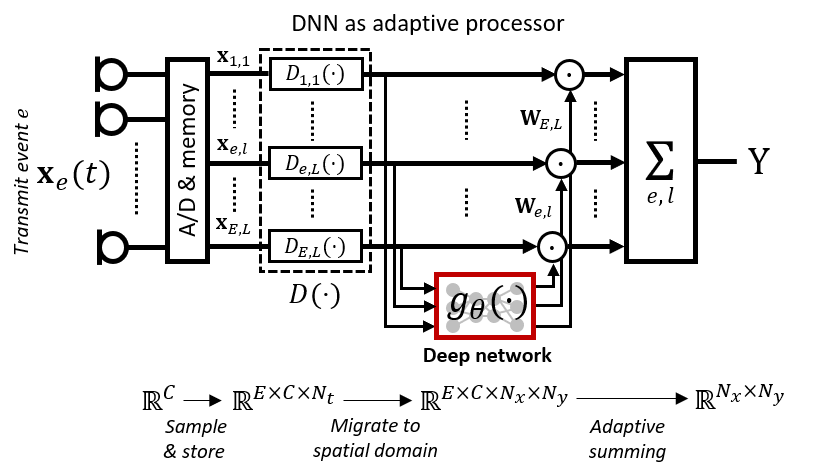}
    \caption{DNN replacing the adaptive processor of classical adaptive beamforming methods.}
    \label{fig:DNN_ABLE}
\end{figure}

The architecture we will discuss now is inspired by the MV beamforming architecture. Instead of replacing the beamforming process entirely, the authors in \cite{luijten2019deep,Luijten2019beamforming} propose to use a deep network as an artificial agent that calculates the optimal apodization weights $\mathrm{W}$ on the fly, given the received pre-delayed channel signals at the array $D(\mathrm{X})$. See Fig.~\ref{fig:DNN_ABLE} for an illustration. By only replacing this bottleneck component in the MVDR beamformer, and constraining the problem further by promoting close-to-distortionless response during training (i.e. $\Sigma_cw_c\approx1$), this solution is highly data-efficient, interpretable, and has the ability to learn powerful models from only few images \cite{Luijten2019beamforming}. \\

\begin{archbox}[DNN as adaptive processor by Luijten et al. ]{arch:luijten}
\noindent\textbf{Acquisition types}: Single plane wave imaging and synthetic aperture (intravascular ultrasound). For the latter, a virtual aperture is constructed by combining the received signals of multiple transmits and receives.\\
\noindent\textbf{Input}: TOF-corrected RF data cube $\mathrm{Z}\in \mathbb{R}^{1 \times C \times N_x \times N_y}$.

\noindent\textbf{Architecture}:
Four convolutional layers comprising 128 nodes for the input and output layers, and 32 nodes for the hidden layers. The kernel size of the filters is $1\times1$, making the receptive field of the network a single pixel. In practice, this is thus a per-pixel fully-connected layer across the array channels. The activation functions are antirectifiers \cite{antirect}, which, unlike ReLUs, preserve both the positive and negative signal components at the expense of a dimensionality increase.

\noindent\textbf{Output}: Array apodization tensor $\mathbf{W}\in\mathbb{R}^{1 \times C \times N_x \times N_y}$, which is subsequently multiplied (element-wise) with the network inputs $\mathrm{Z}$ to yield a beamformed output $\mathrm{Y}$.
\end{archbox}

\noindent\textbf{Complexity, inference speed, and stability} \\
Since pixels are processed independently by the network, a large amount of training data is available per acquisition. Inference is fast and real-time rates are achievable on a GPU-accelerated system. For an array of 128 elements, adaptive calculation of a set of apodization weights through MV beamforming requires $>N^3 (=2,097,152)$ floating point operations (FLOPS), while the deep-learning architecture only requires 74656 FLOPS \cite{Luijten2019beamforming}, in practice leading to a more than $400\times$ speed-up in reconstruction time. Compared to MV beamforming, the deep network is qualitatively more robust, with less observed artefactual reconstructions that stem from e.g. instable computations of the inverse autocorrelation estimates in MV.  %Compared to DAS beamforming, the method provides reduced clutter and enhanced tissue contrast. Quantitatively it yields an elevated contrast-to-noise ratio (10.96~dB vs 11.48~dB), along with significantly improved resolution (0.43~mm vs 0.34~mm, and 0.85~mm vs 0.70~mm in the axial and lateral directions, respectively).

\subsection{DNN for Fourier-domain beamsumming}
\begin{figure}[t!]
    \centering
    \includegraphics[scale=0.7]{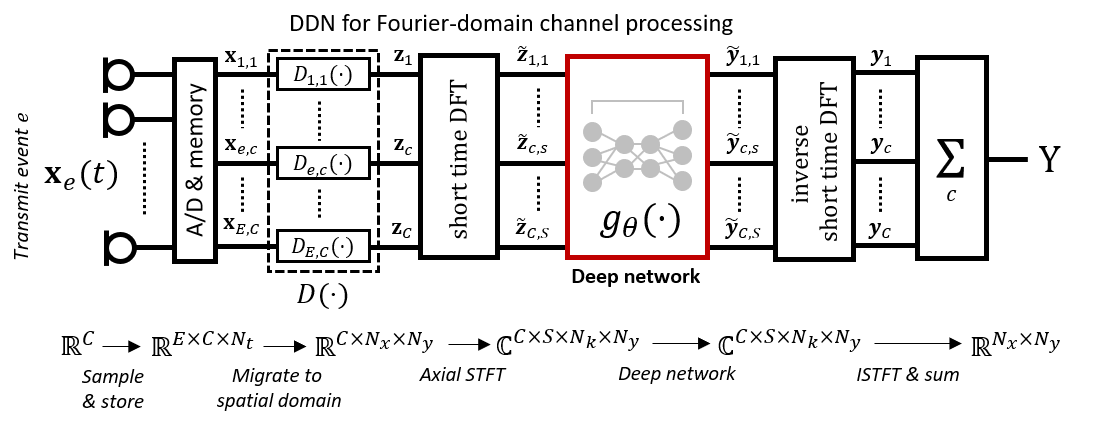}
    \caption{DNN for Fourier-domain beamsumming}
    \label{fig:DNN_STFT}
\end{figure}

Several model-based beamforming methods process ultrasound channel data in the frequency domain \cite{holfort2009broadband}, \cite{byram2015model}. In this spirit, the authors of \cite{luchies2018deep,luchies2019training,luchies2020assessing} use deep networks to perform wideband beamsumming by processing individual DFT bins of the TOF-corrected and axially-windowed RF signals. Each DFT bin is processed using a distinct neural network. After processing in the Fourier domain, the channel signals are summed for each window and a beamformed RF scanline is reconstructed using an inverse short-time Fourier transform (see Fig.~\ref{fig:DNN_STFT}).

\begin{archbox}[DNN for Fourier-domain beamsumming by Luchies et al. ]{arch:luchies}
\noindent\textbf{Acquisition type}: Line scanning, where each transmit events produces one lateral scanline. The first dimension of $\mathrm{X}\in \mathbb{R}^{E \times C \times N_t}$ is thus directly mapped to the lateral dimension $N_y$ in the TOF-correction step: $\mathrm{Z} = D(\mathrm{X}) \in \mathbb{R}^{C \times N_x \times N_y}$.

\noindent\textbf{Input}: Fourier transform of an axially-windowed (window length $S$) and TOF-corrected RF data cube for a single scanline, i.e. $\tilde{\mathrm{Z}}_{n_x, n_y} = \mathcal{F}(\mathrm{Z}_{n_x, n_y}) \in \mathbb{C}^{C \times S \times 1}$, with $\mathrm{Z}_{n_x, n_y}\in \mathbb{R}^{C \times L_x \times 1}$. Before feeding to the network, $\tilde{\mathrm{Z}}_{n_x, n_y}$ is converted to real values by stacking the real and imaginary components, yielding $\tilde{\mathrm{Z}}_{n_x, n_y} \in \mathbb{R}^{2C \times S \times 1}$.

\noindent\textbf{Architecture}:
$S$ identical fully-connected neural networks, one for each DFT bin. Each neural network of this stack thus takes the $2C$ channel values corresponding to that bin as its input. The $S$ networks all have 5 fully-connected layers with the hidden layers having 170 neurons and ReLU activations. Each network then returns $2C$ channel values, the real and imaginary components of that frequency bin after processing.

\noindent\textbf{Output}: Processed Fourier components of the axially-windowed and TOF-corrected RF data cube for a single scanline: $\tilde{\mathrm{Y}}_{n_x,n_y} \in \mathbb{C}^{C \times S \times 1}$. To obtain a beamformed image, the $C$ channels are summed and an inverse short-time Fourier transform is used to compound the responses of all axial windows.

\end{archbox}

\subsection{Post-filtering after beamsumming}
We will now discuss some post-filter approaches, i.e. methods applied after channel beamsumming, but before envelope detection and brightness compression. Several post-filtering methods have been proposed for compounding beamsummed RF outputs from multiple transmit events $e$ with some spatial overlap (e.g. multiple plane/diverging waves). Traditionally, multiple transmit events are compounded by coherent summing (i.e. after transmit delay compensation). Today, deep learning is increasingly used to replace the coherent summing step. In \cite{lu2020reconstruction}, the authors perform neural network compounding from a small number of transmits, and train towards the image obtained by coherently summing a much larger number of transmit events. In \cite{chennakeshava2020high}, the authors pose compounding as an inverse problem, which they subsequently solve using a model-based deep network inspired by proximal gradient methods.
Post-filtering has also been used to e.g. remove aliasing artifacts due to sub-Nyquist sampling on beamformed 1D RF lines \cite{mamistvalov2021deep}. The most common method to perform sparse recovery and solve the L1 minimization problem is using compressed sensing algorithms, such as ISTA and NESTA \cite{eldar2015sampling}. However, they typically suffer from high computational load, and do not always ensure high quality recovery. Here we discuss the process of unfolding an iterative algorithm as the layers of a deep network for sparse recovery. The authors in \cite{mamistvalov2021deep}, built an efficient network by unfolding the ISTA algorithm for sparse recovery, based on the previously suggested LISTA \cite{gregor2010learning}. In their technique, they recover both spatially and temporally sub-Nyquist sampled US data, after delaying it in the frequency domain, using a simple, computationally efficient, and interpretable deep network.

\begin{archbox}[DNN as a recovery method by Mamistvalov et al.]{arch:mamistvalov}
\noindent\textbf{Acquisition type}: Line scanning, so the first dimension (transmit events / lines $E$) has been mapped to the lateral dimension $N_y$ during the time-space migration/TOF correction.

\noindent\textbf{Input}: Frequency domain delayed and summed data (or frequency domain convolutionally beamformed data), after appropriate inverse Fourier transform and appropriate zero padding, to maintain the desired temporal resolution. The input data is a vector, $\mathrm{Z}_{in}\in \mathbb{R}^{N_{st} \times 1}$, where $N_{st}$ is the traditionally used number of samples for beamforming. The recovery is done for each image line separately.

\noindent\textbf{Architecture}:
Simple architecture of unfolded ISTA algorithm, consisting of 30 layers, each includes two convolutional layers that mimic the matrix multiplications of ISTA and one soft thresholding layer. One last convolutional layer is added to recover the actual beamformed signal from the recovered sparse code.

\noindent\textbf{Output}: Beamformed signal for each image line without artifacts caused by sub-Nyquist sampling, $Z_{out}\in \mathbb{R}^{N_{st} \times 1}$.
\end{archbox}

\section{Training strategies and data}
\label{sec:training}

\subsection{Training data}
The model parameters of the above beamforming networks are optimized using training data that consists of simulations, in-vitro/in-vivo data, or a combination thereof. We will now discuss some of the strategies for selecting training data, and in particular generating useful training targets. \\

\noindent \textbf{Simulations} \\
Training ultrasound beamformers using simulated data is appealing, since various ultrasound simulation toolboxes, such as Field II \cite{jensen2004simulation}, k-wave \cite{treeby2010k}, and the Matlab Ultrasound Toolbox (MUST)\footnote{\url{https://www.biomecardio.com/MUST/}", by Damien Garcia},  allow for flexible generation of input-target training data. Simulations can be used to generate pairs of RF data, each pair comprising a realistic imaging mode based on the actual hardware and probe (input), and a second mode of the same scene with various more desirable properties (target). One can get creative with the latter, and we here list a number of popular approaches found in the literature:
\begin{enumerate}
    \item Target imaging mode with unrealistic, yet desired, array and hardware configuration:
    \begin{enumerate}
        \item[a] Higher frequencies/shorter wavelengths (without the increased absorption) to improve target resolution \cite{chennakeshava2020high}.
        \item[b] Larger array aperture to improve target resolution \cite{vignon2020resolution}.
    \end{enumerate}
        \item Removal of undesired imaging effects such as off-axis scattering from target RF data \cite{luchies2019training}.
    \item Targets constructed directly from simulation scene/object:
    \begin{enumerate}
        \item[a] Point targets on a high-resolution simulation grid \cite{youn2020detection}.
        \item[b] Masks of medium properties, e.g. anechoic region segmentations \cite{nair2020deep}.
    \end{enumerate}
\end{enumerate}

When relying solely on simulations, one has to be careful to avoid catastrophic domain shift when deploying the neural models on real data. Increasing the realism of the simulations, mixing simulations with real data, using domain-adaptation methods, or limiting the neural networks receptive field can help combat domain-shift issues. \\

\noindent \textbf{Real data} \\
Training targets from real acquisitions are typically based on high-quality (yet computationally complex) model-based solutions such as MV beamforming or extended/full acquisitions in a compressed sensing setup. In the former, training targets are generated offline by running powerful but time-consuming model-based beamformers on the training data set of RF inputs. The goal of deep-learning-based beamforming is then to achieve the same performance as these model-based solutions, at much faster inference rates. In the compressed sensing setup, training targets are generated by (DAS/MV) beamforming the full (not compressed) set of RF measurements $\mathrm{X}$. In this case, the objective of a deep-learning beamformer is to reproduce these beamformed outputs based on compressed/undersampled measurements  $\mathrm{X}_{u}$. As discussed in Sec.~\ref{sec:opportunities}, compression can entail fast-time sub-Nyquist sampling, imaging with sparse arrays, or limiting the number of transmit events in e.g. plane-wave compounding. Real data is available on the aforementioned CUBDL challenge website \cite{challenge}.

\subsection{Loss functions and optimization}
In this section, we will discuss typical loss functions used to train deep networks for ultrasound beamforming. Most networks are trained by directly optimizing a loss on the beamformer output in the RF or IQ domain. Others indirectly optimize the beamformer output in a task-adaptive fashion by optimizing some downstream loss after additional processing. For training, some variant of stochastic gradient descent (SGD), often Adaptive Moment Estimation (ADAM) is used, and some form of learning rate decay is also common. SGD operates on mini batches, which comprise either full input-output image pairs, or some collection of patches/slices/cubes extracted from full images. In the following, we will (without loss of generality) use $\mathrm{Y}^{(i)}$ and $\mathrm{Y}_{t}^{(i)}$ to refer to respectively the network outputs and targets for a sample $i$. \\

\noindent \textbf{Loss functions for beamformed outputs} \\
Considering image reconstruction as a pixel-wise regression problem under a Gaussian likelihood model, perhaps the most commonly used loss function is the MSE (or $\ell_2$ norm) with respect to the target pixel values:
\begin{equation}
    \mathcal{L}_{MSE} = \frac{1}{I}\sum_{i=0}^{I-1} \left\| \mathrm{Y}^{(i)} -\mathrm{Y}_{t}^{(i)}\right\|_2^2.
    \label{eqn:MSE}
\end{equation}
If one would like to penalize strong deviations less stringently, e.g. to be less sensitive to outliers, one can consider a likelihood model that decays less strongly for large deviations, such as the Laplace distribution. Under that model, the negative log likelihood loss function is the mean absolute error (or $\ell_1$ norm):
\begin{equation}
    \mathcal{L}_{l1} = \frac{1}{I}\sum_{i=0}^{I-1} \left\| \mathrm{Y}^{(i)} -\mathrm{Y}_{t}^{(i)}\right\|_1.
    \label{eqn:l1}
\end{equation}

A commonly adopted variant of the MSE loss is the signed-mean-squared-logarithmic-error SMSLE, proposed by Luijten et al. \cite{luijten2019deep}. This metric compresses the large dynamic range of backscattered ultrasound RF signals to promote accurate reconstructions across the entire dynamic range:
\begin{align}
    \mathcal{L}_{SMSLE} =& \frac{1}{2I}\sum_{i=0}^{I-1} \left\| \log_{10}\left(\mathrm{Y}^{(i)}\right)^+ - \log_{10}\left(\mathrm{Y}_{t}^{(i)}\right)^+\right\|_2^2  \\
    +& \left\| \log_{10}\left(\mathrm{Y}^{(i)}\right)^- - \log_{10}\left(\mathrm{Y}_{t}^{(i)}\right)^- \right\|_2^2
    \label{eqn:SMSLE},
\end{align}
where $(\cdot)^+$ and $(\cdot)^-$ yield the magnitude of the positive and negative parts, respectively.
Thus far, we have only covered pixel-wise losses that consider every pixel as an independent sample. These losses have no notion of spatial context and do not measure structural deviations. The structural similarity index (SSIM) \cite{wang2004image} aims to quantify perceived change in structural information,  luminance and contrast. In the vein of the SMSLE, Kessler et al. \cite{kessler2020deep} propose a SSIM loss for ultrasound beamforming that acts on the log-compressed positive and negative parts of the beamformed RF signals:
\begin{align}
    \mathcal{L}_{SSIM} =& \frac{1}{2I}\sum_{i=0}^{I-1}  \left(1-SSIM\left(\log_{10}\left(\mathrm{Y}^{(i)}\right)^+, \log_{10}\left(\mathrm{Y}_{t}^{(i)}\right)^+\right)\right)  \\
    +& \left(1-SSIM\left(\log_{10}\left(\mathrm{Y}^{(i)}\right)^-, \log_{10}\left(\mathrm{Y}_{t}^{(i)}\right)^-\right)\right)
    \label{eqn:SSIM},
\end{align}
where, when luminance, contrast and structure are weighed equally, $SSIM$ is defined as:
\begin{equation}
    SSIM(a,b) = \frac{(2\mu_a\mu_b+\epsilon_1)(2\sigma_{ab}+\epsilon_2)}{(\mu_a^2+\mu_b^2+\epsilon_1)(\sigma_a^2+\sigma_b^2+\epsilon_2)},
\end{equation}
with $\mu_a$, $\mu_b$, $\sigma_a$, $\sigma_b$, and $\sigma_{ab}$ being the means, standard deviations and cross-correlation of $a$ and $b$, and $\epsilon_1$, $\epsilon_2$ being small constants to stabilize the division.

Beyond distance measurements between pixel values, some authors make use of specific adversarial optimization schemes that aim to match the distributions of the targets and generated outputs \cite{chennakeshava2020high}. These approaches make use of a second neural network, the adversary or discriminator, that is trained to discriminate between images that are drawn from the distribution of targets, and those that are generated by the beamforming neural network. This is achieved by minimizing the binary cross-entropy classification loss between its predictions and the labels (target or generated), evaluated on batches that contain both target images and generated images. At the same time, the beamforming network is trained to maximize this loss, thereby attempting to fool the discriminator. The rationale here is that the probability distributions of target images and beamformed network outputs match (or strongly overlap) whenever this neural discriminator cannot distinguish images from either distribution anymore. The beamforming network and discriminator thus play a min-max game, expressed by the following optimization problem across the training data distribution $P_\mathcal{D}$:
\begin{equation}
\label{eqn:optimization_adversary}
\hat{\theta},\hat{\bm{\Psi}} =  \underset{\mathbb{\psi}}{\mathrm{argmin}}~ \underset{\mathbb{\theta}}{\mathrm{argmax}}
\Bigr\{-\mathbb{E}_{(\mathrm{X},\mathrm{Y}_t)\sim P_\mathcal{D}}\left[
        \log(D_{\psi}(\mathrm{Y}_t))+\log(1-D_{\psi}(f_\theta(\mathrm{X})))
        \right]\Bigr\},
\end{equation}
where $\theta$ are the parameters of the beamforming network $f_\theta(\cdot)$ and $\psi$ are the parameters of the discriminator $D_\psi$.
It is important to realize that merely matching distributions does not guarantee accurate image reconstructions. That is why adversarial losses are often applied on input-output and input-target pairs (matching e.g. their joint distributions), or used in combination with additional distance metrics such as those discussed earlier in this section. The relative contributions or weighting of these individual loss terms is typically selected empirically. \\

\noindent \textbf{Task-adaptive optimization} \\
As discussed in Sec.~\ref{sec:opportunities}, one can also optimize the parameters of beamforming architectures using a downstream task-based loss, i.e.:
\begin{equation}
\label{eqn:optimization_task}
\hat{\theta},\hat{\bm{\Phi}} =  \underset{\mathbb{\theta}}{\mathrm{argmin}}
\Bigr\{\mathbb{E}_{(\mathrm{X}, s_{task})\sim P_\mathcal{D}}\left[\mathcal{L}_{task}
\Bigr\lbrace
g_\phi\left(f_\theta(\mathrm{X})\right),s_{task}
\Bigr\rbrace
\right]
\Bigr\},
\end{equation}
where $s_t$ denotes some target task, $\mathcal{L}_{task}\{a,b\}$ is a task-specific loss function between outputs $a$ and targets $b$, and $\phi$ are the the parameters of the task function $g_\psi$, which can be a neural network. Examples of such tasks include segmentation \cite{nair2020deep}, for which $\mathcal{L}_{task}$ is e.g. a Dice loss, or motion estimation (Doppler), for which $\mathcal{L}_{task}$ is e.g. an MSE penalty \cite{huijben2020learning}.

\section{New Research Opportunities}
\label{sec:new}

\subsection{Multi-functional deep beamformer}

Although deep beamfomers provide impressive
performance and ultra-fast reconstruction, one of the downsides of the deep beamformers is that
a distinct model is needed for each type of desired output. For instance, to obtain DAS outputs, a model is needed which mimics DAS; similarly, for MVBF a separate model is needed. Although the architecture of the model could be the same,  separate weights need to be stored for each output type.
Given that hundreds/thousands of B-mode optimizations/settings are used in the current high-end commercial systems, one may wonder whether we need to store thousands of deep models in the scanner to deal with various B-mode settings.

\begin{figure}[!hbt]
	\centerline{\includegraphics[width=1\columnwidth]{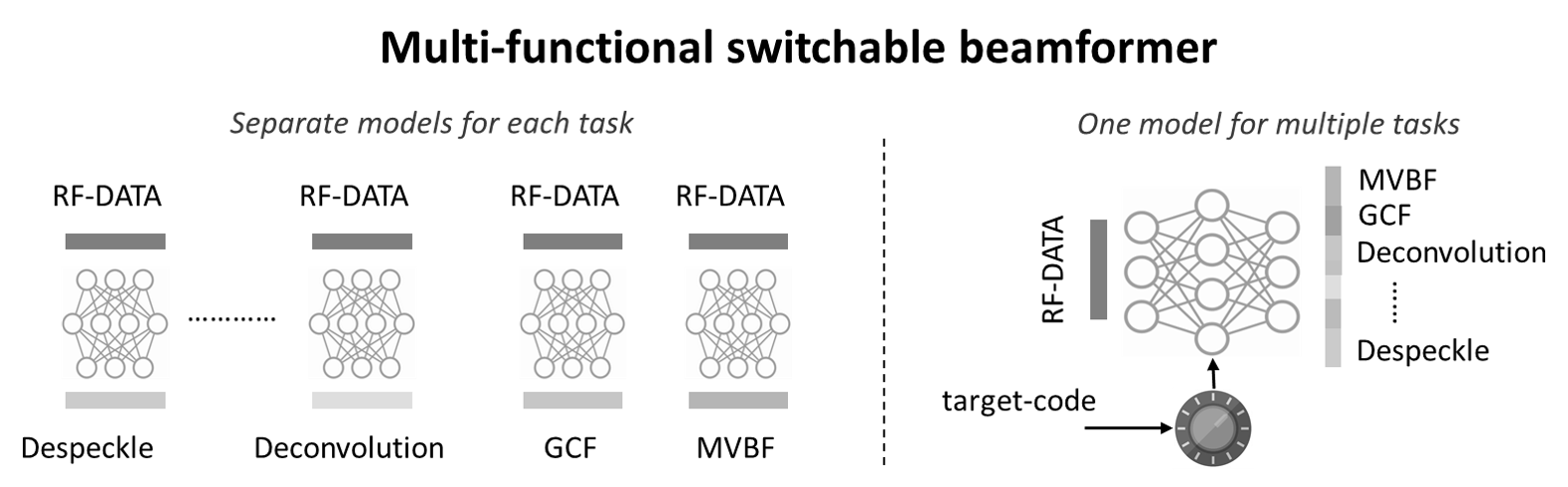}}
    \caption{An illustration of switchable deep beamformer using AdaIN layer.}
    \label{fig:AdaIN}
\end{figure}

To address this issue, \cite{khan2020switchable} recently proposed a {\em switchable} deep beamformer architecture using adaptive instance normalization (AdaIN) layers as shown in Fig.~\ref{fig:AdaIN}.
Specifically,
AdaIN was originally proposed as an image style transfer method,
in which the mean and variance of the feature vectors are replaced by those of the style reference image \cite{huang2017arbitrary}.
Suppose that a multi-channel feature tensor at a specific layer is represented by
\begin{align}\label{eq:X}
\Xb =&\begin{bmatrix} \xb_1  & \cdots &\xb_C \end{bmatrix} \in \Rd^{HW\times C},
\end{align}
where $C$ is  the number of channels in the feature tensor $\Xb$, and
$\xb_i  \in \Rd^{HW\times 1}$ refers to the $i$-th column vector of $\Xb$,
which represents the vectorized feature map of size of $H\times W$  at the $i$-th channel.
%Suppose, furthermore, the corresponding feature map for the style reference image is given by
% \begin{align}\label{eq:Y}
%\yb =&\begin{bmatrix} \yb_1  & \cdots &\yb_P \end{bmatrix} \in \Rd^{HW\times P} .
%\end{align}
Then,  AdaIN \cite{huang2017arbitrary} converts the feature data at each channel using the following transform:
\begin{align}
\zb_i &= \Tc(\xb_i,\yb_i),\quad i=1,\cdots, C
\end{align}
where
\begin{align}\label{eq:AdaIN}
\Tc(\xb,\yb)  := \frac{\sigma(\yb) }{\sigma(\xb) }\left(\xb -m(\xb) \1\right) +m(\yb)\1,\quad
\end{align}
where $\1 \in \Rd^{HW}$ is the $HW$-dimensional vector composed of 1, and
 $m(\xb)$ and $\sigma(\xb)$ are the mean and standard deviation of $\xb\in \Rd^{HW}$;
$m(\yb)$ and $\sigma(\yb)$ refer to the target style domain mean and standard deviation, respectively.
Eq.~\eqref{eq:AdaIN} implies that the mean and variance of the feature in the input image are
 normalized so that they can match the mean and variance of the style image feature.
Although \eqref{eq:AdaIN} looks heuristic, it was shown that  the transform \eqref{eq:AdaIN} is closely related to the optimal transport between two Gaussian probability distributions \cite{peyre2019computational,villani2008optimal}.

Inspired by this,  \cite{khan2020switchable} demonstrate that
a {\em single} deep beamformer with AdaIN layers can learn target images from various styles. % for training.
Here, a ``style'' refers to a specific output processing, such as DAS, MVBF, deconvolution image, despeckled images, etc.
Once the network is trained,
the deep beamformer can then generate various style output
by simply changing the AdaIN code.
Furthermore, the AdaIN code generation is easily performed with a very light  AdaIN code generator,
so the additional memory overhead at the training step is minimal.
Once the neural network is trained, we only need the AdaIN codes without the generator,
which makes the system even simpler.

\subsection{Unsupervised Learning}
As discussed before, most existing deep learning strategies for ultrasound beamforming are based on supervised learning, thus relying predominantly on paired input-target datasets. However, in many real world imaging situations, access to paired images (input channel data and a corresponding desired output image) is not possible. For example, to improve the visual quality of US images acquired using a low-cost imaging system we need to scan exactly the same field of view using a high-end machine, which is not trivial. For denoising or artifact removal, the actual ground-truth is not known in-vivo, so supervised learning approaches are typically left with simulation datasets for training. This challenge has spurred a growing interest in developing an unsupervised learning strategy where channel data from low-end system or artifact corruption can be used as inputs, using surrogate performance metrics (based on high quality images from different machines and imaging conditions, or statistical properties) to train networks.
		
One possible approach to address this problem is to adopt unpaired style transfer strategies based on e.g. CycleGANs - a technique that has shown successful for many image domain quality improvements, also in ultrasound. For example, the authors in \cite{jafari2020cardiac,khan2021variational} employed such a cycleGAN to improve the image quality from portable US image using high-end unmatched image data. In general, approaches that drive training by matching distribution properties (e.g. through discriminator networks as in CycleGAN) rather than strict input-output pairs hold promise for such applications.

\bibliography{percolation,main}\label{refs}

@article{challengepaper,
  title={Deep Learning for Ultrasound Image Formation: CUBDL Evaluation Framework and Open Datasets},
  author={Hyun, D. and Wiacek, A. and Goudarzi, S. and Rothlübbers, S. and Asif, A. and Eickel, K. and Eldar, Y.C. and Huang, J. and Mischi, M. and Rivaz, H. and Sinden, D. and van Sloun, R.J.G. and Strohm, H. and Bell, M.A.L.},
  journal={IEEE Transactions on Ultrasonics, Ferroelectrics and Frequency Control (accepted)},
  year={2021}
}

@article{mamistvalov2020compressed,
  title={Compressed Fourier-Domain Convolutional Beamforming for Wireless Ultrasound imaging},
  author={Mamistvalov, Alon and Eldar, Yonina C},
  journal={arXiv preprint arXiv:2010.13171},
  year={2020}
}

@article{cohen2020sparse,
  title={Sparse array design via fractal geometries},
  author={Cohen, Regev and Eldar, Yonina C},
  journal={IEEE transactions on signal processing},
  volume={68},
  pages={4797--4812},
  year={2020},
  publisher={IEEE}
}

@misc{challenge,
  author = {Bell, Muyinatu and Huang, Jiaqi and Hyung, Dongwoon and Eldar, Yonina and van Sloun, Ruud and Mischi, Massimo},
  title = {{Challenge on Ultrasound Beamforming by Deep Learning}},
  howpublished = "\url{cubdl.jhu.edu}",
  year = {2020},
  note = "[Online; accessed 15-September-2021]"
}

@article{asl2010eigenspace,
  title={Eigenspace-based minimum variance beamforming applied to medical ultrasound imaging},
  author={Asl, Babak Mohammadzadeh and Mahloojifar, Ali},
  journal={IEEE transactions on ultrasonics, ferroelectrics, and frequency control},
  volume={57},
  number={11},
  pages={2381--2390},
  year={2010},
  publisher={IEEE}
}

@inproceedings{huang2017arbitrary,
  title={Arbitrary style transfer in real-time with adaptive instance normalization},
  author={Huang, Xun and Belongie, Serge},
  booktitle={Proceedings of the IEEE International Conference on Computer Vision},
  pages={1501--1510},
  year={2017}
}

@article{jafari2020cardiac,
  title={Cardiac point-of-care to cart-based ultrasound translation using constrained CycleGAN},
  author={Jafari, Mohammad H and Girgis, Hany and Van Woudenberg, Nathan and Moulson, Nathaniel and Luong, Christina and Fung, Andrea and Balthazaar, Shane and Jue, John and Tsang, Micheal and Nair, Parvathy and others},
  journal={International journal of computer assisted radiology and surgery},
  volume={15},
  number={5},
  pages={877--886},
  year={2020},
  publisher={Springer}
}

@ARTICLE{khan2021variational,
  author={S. {Khan} and J. {Huh} and J. C. {Ye}},
  journal={IEEE Transactions on Ultrasonics, Ferroelectrics, and Frequency Control},
  title={Variational Formulation of Unsupervised Deep Learning for Ultrasound Image Artifact Removal},
  year={2021},
  volume={},
  number={},
  pages={1-1},
  doi={10.1109/TUFFC.2021.3056197}}

@book{villani2008optimal,
  title={Optimal transport: old and new},
  author={Villani, C{\'e}dric},
  volume={338},
  year={2008},
  publisher={Springer Science \& Business Media}
}

@article{peyre2019computational,
  title={Computational Optimal Transport: With Applications to Data Science},
  author={Peyr{\'e}, Gabriel and Cuturi, Marco and others},
  journal={Foundations and Trends{\textregistered} in Machine Learning},
  volume={11},
  number={5-6},
  pages={355--607},
  year={2019},
  publisher={Now Publishers, Inc.}
}

@string{icassp = {Proc. of IEEE Int'l Conf. on Acoust., Speech and Sig. Proc.}}

@book{eldar2015sampling,
  title={Sampling theory: Beyond bandlimited systems},
  author={Eldar, Yonina C},
  year={2015},
  publisher={Cambridge University Press}
}

@article{christensen2020super,
  title={Super-resolution ultrasound imaging},
  author={Christensen-Jeffries, Kirsten and Couture, Olivier and Dayton, Paul A and Eldar, Yonina C and Hynynen, Kullervo and Kiessling, Fabian and O'Reilly, Meaghan and Pinton, Gianmarco F and Schmitz, Georg and Tang, Meng-Xing and others},
  journal={Ultrasound in medicine \& biology},
  volume={46},
  number={4},
  pages={865--891},
  year={2020},
  publisher={Elsevier}
}

@article{treeby2010k,
  title={k-Wave: MATLAB toolbox for the simulation and reconstruction of photoacoustic wave fields},
  author={Treeby, Bradley E and Cox, Benjamin T},
  journal={Journal of biomedical optics},
  volume={15},
  number={2},
  pages={021314},
  year={2010},
  publisher={International Society for Optics and Photonics}
}

@inproceedings{jensen2004simulation,
  title={Simulation of advanced ultrasound systems using Field II},
  author={Jensen, J{\o}rgen Arendt},
  booktitle={2004 2nd IEEE International Symposium on Biomedical Imaging: Nano to Macro (IEEE Cat No. 04EX821)},
  pages={636--639},
  year={2004},
  organization={IEEE}
}

@article{youn2020detection,
  title={Detection and localization of ultrasound scatterers using convolutional neural networks},
  author={Youn, Jihwan and Ommen, Martin Lind and Stuart, Matthias Bo and Thomsen, Erik Vilain and Larsen, Niels Bent and Jensen, J{\o}rgen Arendt},
  journal={IEEE Transactions on Medical Imaging},
  volume={39},
  number={12},
  pages={3855--3867},
  year={2020},
  publisher={IEEE}
}

@inproceedings{vignon2020resolution,
  title={Resolution Improvement with a Fully Convolutional Neural Network Applied to Aligned Per-Channel data},
  author={Vignon, Francois and Shin, Jun Seob and Meral, Faik Can and Apostolakis, Iason and Huang, Sheng-Wen and Robert, Jean-Luc},
  booktitle={2020 IEEE International Ultrasonics Symposium (IUS)},
  pages={1--4},
  year={2020},
  organization={IEEE}
}

@article{byram2015model,
  title={A model and regularization scheme for ultrasonic beamforming clutter reduction},
  author={Byram, Brett and Dei, Kazuyuki and Tierney, Jaime and Dumont, Douglas},
  journal={IEEE transactions on ultrasonics, ferroelectrics, and frequency control},
  volume={62},
  number={11},
  pages={1913--1927},
  year={2015},
  publisher={IEEE}
}

@article{holfort2009broadband,
  title={Broadband minimum variance beamforming for ultrasound imaging},
  author={Holfort, Iben Kraglund and Gran, Fredrik and Jensen, Jorgen Arendt},
  journal={IEEE transactions on ultrasonics, ferroelectrics, and frequency control},
  volume={56},
  number={2},
  pages={314--325},
  year={2009},
  publisher={IEEE}
}

@misc{antirect,
  title = {Antirectifier},
  author = {Fran{\c{c}}ois Chollet},
  howpublished = {\url{https://github.com/keras-team/keras/blob/master/examples/antirectifier.py}},
  note = {Accessed: 14-03-2021}
}

@article{wang2004image,
  title={Image quality assessment: from error visibility to structural similarity},
  author={Wang, Zhou and Bovik, Alan C and Sheikh, Hamid R and Simoncelli, Eero P},
  journal={IEEE transactions on image processing},
  volume={13},
  number={4},
  pages={600--612},
  year={2004},
  publisher={IEEE}
}

@article{lu2020reconstruction,
  title={Reconstruction for diverging-wave imaging using deep convolutional neural networks},
  author={Lu, Jingfeng and Millioz, Fabien and Garcia, Damien and Salles, S{\'e}bastien and Liu, Wanyu and Friboulet, Denis},
  journal={IEEE transactions on ultrasonics, ferroelectrics, and frequency control},
  volume={67},
  number={12},
  pages={2481--2492},
  year={2020},
  publisher={IEEE}
}

@inproceedings{chennakeshava2020high,
  title={High Resolution Plane Wave Compounding Through Deep Proximal Learning},
  author={Chennakeshava, Nishith and Luijten, Ben and Drori, Oded and Mischi, Massimo and Eldar, Yonina C and van Sloun, Ruud JG},
  booktitle={2020 IEEE International Ultrasonics Symposium (IUS)},
  pages={1--4},
  year={2020},
  organization={IEEE}
}

@article{mamistvalov2021deep,
  title={Deep Unfolded Recovery of Sub-Nyquist Sampled Ultrasound Image},
  author={Mamistvalov, Alon and Eldar, Yonina C},
  journal={arXiv preprint arXiv:2103.01263},
  year={2021}
}

@article{nair2020deep,
  title={Deep learning to obtain simultaneous image and segmentation outputs from a single input of raw ultrasound channel data},
  author={Nair, Arun Asokan and Washington, Kendra N and Tran, Trac D and Reiter, Austin and Bell, Muyinatu A Lediju},
  journal={IEEE transactions on ultrasonics, ferroelectrics, and frequency control},
  volume={67},
  number={12},
  pages={2493--2509},
  year={2020},
  publisher={IEEE}
}

@article{wiacek2020coherenet,
  title={Coherenet: A deep learning architecture for ultrasound spatial correlation estimation and coherence-based beamforming},
  author={Wiacek, Alycen and Gonz{\'a}lez, Eduardo and Bell, Muyinatu A Lediju},
  journal={IEEE transactions on ultrasonics, ferroelectrics, and frequency control},
  volume={67},
  number={12},
  pages={2574--2583},
  year={2020},
  publisher={IEEE}
}

@article{kessler2020deep,
  title={Deep-Learning Based Adaptive Ultrasound Imaging from Sub-Nyquist Channel Data},
  author={Kessler, Naama and Eldar, Yonina C},
  journal={arXiv preprint arXiv:2008.02628},
  year={2020}
}

@article{luchies2020assessing,
  title={Assessing the Robustness of Frequency-Domain Ultrasound Beamforming Using Deep Neural Networks},
  author={Luchies, Adam C and Byram, Brett C},
  journal={IEEE Transactions on Ultrasonics, Ferroelectrics, and Frequency Control},
  volume={67},
  number={11},
  pages={2321--2335},
  year={2020},
  publisher={IEEE}
}

@article{luchies2019training,
  title={Training improvements for ultrasound beamforming with deep neural networks},
  author={Luchies, Adam C and Byram, Brett C},
  journal={Physics in Medicine \& Biology},
  volume={64},
  number={4},
  pages={045018},
  year={2019},
  publisher={IOP Publishing}
}

@article{luchies2018deep,
  title={Deep neural networks for ultrasound beamforming},
  author={Luchies, Adam C and Byram, Brett C},
  journal={IEEE transactions on medical imaging},
  volume={37},
  number={9},
  pages={2010--2021},
  year={2018},
  publisher={IEEE}
}

@article{khan2020switchable,
  title={Switchable Deep Beamformer},
  author={Khan, Shujaat and Huh, Jaeyoung and Ye, Jong Chul},
  journal={arXiv preprint arXiv:2008.13646},
  year={2020}
}

@article{khan2020adaptive,
  title={Adaptive and compressive beamforming using deep learning for medical ultrasound},
  author={Khan, Shujaat and Huh, Jaeyoung and Ye, Jong Chul},
  journal={IEEE transactions on ultrasonics, ferroelectrics, and frequency control},
  volume={67},
  number={8},
  pages={1558--1572},
  year={2020},
  publisher={IEEE}
}

@article{yoon2018efficient,
  title={Efficient b-mode ultrasound image reconstruction from sub-sampled rf data using deep learning},
  author={Yoon, Yeo Hun and Khan, Shujaat and Huh, Jaeyoung and Ye, Jong Chul},
  journal={IEEE transactions on medical imaging},
  volume={38},
  number={2},
  pages={325--336},
  year={2018},
  publisher={IEEE}
}

@book{sutton2018reinforcement,
  title={Reinforcement learning: An introduction},
  author={Sutton, Richard S and Barto, Andrew G},
  year={2018},
  publisher={MIT press}
}

@inproceedings{kulkarni2016reconnet,
  title={Reconnet: Non-iterative reconstruction of images from compressively sensed measurements},
  author={Kulkarni, Kuldeep and Lohit, Suhas and Turaga, Pavan and Kerviche, Ronan and Ashok, Amit},
  booktitle={Proceedings of the IEEE Conference on Computer Vision and Pattern Recognition},
  pages={449--458},
  year={2016}
}

@article{williams1992simple,
  title={Simple statistical gradient-following algorithms for connectionist reinforcement learning},
  author={Williams, Ronald J},
  journal={Machine learning},
  volume={8},
  number={3-4},
  pages={229--256},
  year={1992},
  publisher={Springer}
}

@article{huijben2020learning,
  title={Learning sub-sampling and signal recovery with applications in ultrasound imaging},
  author={Huijben, Iris AM and Veeling, Bastiaan S and Janse, Kees and Mischi, Massimo and van Sloun, Ruud JG},
  journal={IEEE Transactions on Medical Imaging},
  volume={39},
  number={12},
  pages={3955--3966},
  year={2020},
  publisher={IEEE}
}

@article{lorintiu2015compressed,
  title={Compressed sensing reconstruction of {3D} ultrasound data using dictionary learning and line-wise subsampling},
  author={Lorintiu, Oana and Liebgott, Herv{\'e} and Alessandrini, Martino and Bernard, Olivier and Friboulet, Denis},
  journal={IEEE Trans. Med. Imag.},
  volume={34},
  number={12},
  pages={2467--2477},
  year={2015},
  publisher={IEEE}
}

@article{austeng2002sparse,
  title={Sparse 2-D arrays for 3-D phased array imaging-design methods},
  author={Austeng, Andreas and Holm, Sverre},
  journal={IEEE Trans. Ultrason., Ferroelectr., Freq. Control},
  volume={49},
  number={8},
  pages={1073--1086},
  year={2002},
  publisher={IEEE}
}

@inproceedings{liu2017maximally,
  title={Maximally economic sparse arrays and Cantor arrays},
  author={Liu, Chun-Lin and Vaidyanathan, PP},
  booktitle={2017 IEEE 7th Int. Workshop on Computational Advances in Multi-Sensor Adaptive Processing (CAMSAP)},
  pages={1--5},
  year={2017},
  organization={IEEE}
}

@inproceedings{besson2016compressed_1,
  title={A compressed beamforming framework for ultrafast ultrasound imaging},
  author={Besson, Adrien and Carrillo, Rafael E and Perdios, Dimitris and Arditi, Marcel and Bernard, Olivier and Wiaux, Yves and Thiran, Jean-Philippe},
  booktitle={2016 IEEE Int. Ultrasonics Symposium (IUS)},
  pages={1--4},
  year={2016},
  organization={IEEE}
}

@article{cohen2018beamforming,
  title={Sparse Convolutional Beamforming for Ultrasound Imaging},
  author={Cohen, Regev and Eldar, Yonina C},
  journal={IEEE Trans. Ultrason., Ferroelectr., Freq. Control},
  volume={65},
  number={12},
  pages={2390--2406},
  year={2018},
  publisher={IEEE}
}

@article{cohen2018sparse,
  title={Sparse Doppler sensing based on nested arrays},
  author={Cohen, Regev and Eldar, Yonina C},
  journal={IEEE Trans. Ultrason., Ferroelectr., Freq. Control},
  volume={65},
  number={12},
  pages={2349--2364},
  year={2018},
  publisher={IEEE}
}

@article{nilsen2010wiener,
  title={Wiener beamforming and the coherence factor in ultrasound imaging},
  author={Nilsen, Carl-Inge Colombo and Holm, Sverre},
  journal={IEEE transactions on ultrasonics, ferroelectrics, and frequency control},
  volume={57},
  number={6},
  pages={1329--1346},
  year={2010},
  publisher={IEEE}
}

@article{vanSloun2019DL_in_US,
  title={Deep learning in ultrasound imaging},
  author={Van Sloun, Ruud JG and Cohen, Regev and Eldar, Yonina C},
  journal={Proceedings of the IEEE},
  volume={108},
  number={1},
  pages={11--29},
  year={2019},
  publisher={IEEE}
}

@inproceedings{luijten2019deep,
  title={Deep Learning for Fast Adaptive Beamforming},
  author={Luijten, Ben and Cohen, Regev and de Bruijn, Frederik J and Schmeitz, Harold AW and Mischi, Massimo and Eldar, Yonina C and van Sloun, Ruud JG},
  booktitle={ICASSP 2019-2019 IEEE International Conference on Acoustics, Speech and Signal Processing (ICASSP)},
  pages={1333--1337},
  year={2019},
  organization={IEEE}
}

@ARTICLE{Bae2016,
author={M. {Bae} and S. B. {Park} and S. J. {Kwon}},
journal={IEEE Transactions on Ultrasonics, Ferroelectrics, and Frequency Control},
title={Fast Minimum Variance Beamforming Based on Legendre Polynomials},
year={2016},
volume={63},
number={9},
pages={1422-1431},
doi={10.1109/TUFFC.2016.2591623},
ISSN={0885-3010},
month={Sep.},}

@ARTICLE{Kim2014,
author={K. {Kim} and S. {Park} and J. {Kim} and S. {Park} and M. {Bae}},
journal={IEEE Transactions on Ultrasonics, Ferroelectrics, and Frequency Control},
title={A fast minimum variance beamforming method using principal component analysis},
year={2014},
volume={61},
number={6},
pages={930-945},
doi={10.1109/TUFFC.2014.2989},
ISSN={0885-3010},
month={June},}

@article{Mallart1994,
author = {Mallart,Raoul  and Fink,Mathias },
title = {Adaptive focusing in scattering media through sound‐speed inhomogeneities: The van Cittert Zernike approach and focusing criterion},
journal = {The Journal of the Acoustical Society of America},
volume = {96},
number = {6},
pages = {3721-3732},}

@ARTICLE{Chernyakova2019,
author={T. {Chernyakova} and D. {Cohen} and M. {Shoham} and Y. C. {Eldar}},
journal={IEEE Transactions on Ultrasonics, Ferroelectrics, and Frequency Control},
title={iMAP Beamforming for High Quality High Frame Rate Imaging},
year={2019},
volume={},
number={},
pages={1-1},
doi={10.1109/TUFFC.2019.2933506},
ISSN={0885-3010},
month={},}

@ARTICLE{Nilsen2010,
author={C. C. {Nilsen} and S. {Holm}},
journal={IEEE Transactions on Ultrasonics, Ferroelectrics, and Frequency Control},
title={Wiener beamforming and the coherence factor in ultrasound imaging},
year={2010},
volume={57},
number={6},
pages={1329-1346},
doi={10.1109/TUFFC.2010.1553},
ISSN={0885-3010},
month={June},}

@InProceedings{Khan2019,
author="Khan, Shujaat
and Huh, Jaeyoung
and Ye, Jong Chul",
editor="Shen, Dinggang
and Liu, Tianming
and Peters, Terry M.
and Staib, Lawrence H.
and Essert, Caroline
and Zhou, Sean
and Yap, Pew-Thian
and Khan, Ali",
title="Deep Learning-Based Universal Beamformer for Ultrasound Imaging",
booktitle="Medical Image Computing and Computer Assisted Intervention -- MICCAI 2019",
year="2019",
publisher="Springer International Publishing",
address="Cham",
pages="619--627",
isbn="978-3-030-32254-0"
}

@article{hyun2019beamforming,
  title={Beamforming and Speckle Reduction Using Neural Networks},
  author={Hyun, Dongwoon and Brickson, Leandra L and Looby, Kevin T and Dahl, Jeremy J},
  journal={IEEE transactions on ultrasonics, ferroelectrics, and frequency control},
  year={2019},
  publisher={IEEE}
}

@article{wagner2012compressed,
  title={Compressed beamforming in ultrasound imaging},
  author={Wagner, Noam and Eldar, Yonina C and Friedman, Zvi},
  journal={IEEE Transactions on Signal Processing},
  volume={60},
  number={9},
  pages={4643--4657},
  year={2012},
  publisher={IEEE}
}

@article{lahav2017focus,
  title={FoCUS: Fourier-based coded ultrasound},
  author={Lahav, Almog and Chernyakova, Tanya and Eldar, Yonina C},
  journal={IEEE transactions on ultrasonics, ferroelectrics, and frequency control},
  volume={64},
  number={12},
  pages={1828--1839},
  year={2017},
  publisher={IEEE}
}

@article{chernyakova2018fourier,
  title={Fourier-domain beamforming and structure-based reconstruction for plane-wave imaging},
  author={Chernyakova, Tanya and Cohen, Regev and Mulayoff, Rotem and Sde-Chen, Yael and Fraschini, Christophe and Bercoff, Jeremy and Eldar, Yonina C},
  journal={IEEE transactions on ultrasonics, ferroelectrics, and frequency control},
  volume={65},
  number={10},
  pages={1810--1821},
  year={2018},
  publisher={IEEE}
}

@article{burshtein2016sub,
  title={Sub-Nyquist sampling and Fourier domain beamforming in volumetric ultrasound imaging},
  author={Burshtein, Amir and Birk, Michael and Chernyakova, Tanya and Eilam, Alon and Kempinski, Arcady and Eldar, Yonina C},
  journal={IEEE transactions on ultrasonics, ferroelectrics, and frequency control},
  volume={63},
  number={5},
  pages={703--716},
  year={2016},
  publisher={IEEE}
}

@article{demene2015spatiotemporal,
  title={Spatiotemporal clutter filtering of ultrafast ultrasound data highly increases Doppler and fUltrasound sensitivity},
  author={Demen{\'e}, Charlie and Deffieux, Thomas and Pernot, Mathieu and Osmanski, Bruno-F{\'e}lix and Biran, Val{\'e}rie and Gennisson, Jean-Luc and Sieu, Lim-Anna and Bergel, Antoine and Franqui, Stephanie and Correas, Jean-Michel and others},
  journal={IEEE transactions on medical imaging},
  volume={34},
  number={11},
  pages={2271--2285},
  year={2015},
  publisher={IEEE}
}

@article{provost20143d,
  title={{3D} ultrafast ultrasound imaging in vivo},
  author={Provost, Jean and Papadacci, Clement and Arango, Juan Esteban and Imbault, Marion and Fink, Mathias and Gennisson, Jean-Luc and Tanter, Mickael and Pernot, Mathieu},
  journal={Physics in Medicine \& Biology},
  volume={59},
  number={19},
  pages={L1},
  year={2014},
  publisher={IOP Publishing}
}

@article{tanter2014ultrafast,
  title={Ultrafast imaging in biomedical ultrasound},
  author={Tanter, Mickael and Fink, Mathias},
  journal={IEEE transactions on ultrasonics, ferroelectrics, and frequency control},
  volume={61},
  number={1},
  pages={102--119},
  year={2014},
  publisher={IEEE}
}

@inproceedings{baran2009design,
  title={Design of low-cost portable ultrasound systems},
  author={Baran, Jonathan M and Webster, John G},
  booktitle={Annual International Conference of the IEEE Engineering in Medicine and Biology Society},
  pages={792--795},
  year={2009},
  organization={IEEE}
}

@book{szabo2004diagnostic,
  title={Diagnostic ultrasound imaging: inside out},
  author={Szabo, Thomas L},
  year={2004},
  publisher={Academic Press}
}

@article{bercoff2004supersonic,
  title={Supersonic shear imaging: a new technique for soft tissue elasticity mapping},
  author={Bercoff, J{\'e}r{\'e}my and Tanter, Mickael and Fink, Mathias},
  journal={IEEE transactions on ultrasonics, ferroelectrics, and frequency control},
  volume={51},
  number={4},
  pages={396--409},
  year={2004},
  publisher={IEEE}
}

@inproceedings{wagner2011xampling,
  title={Xampling in ultrasound imaging},
  author={Wagner, Noam and Eldar, Yonina C and Feuer, Arie and Danin, Gilad and Friedman, Zvi},
  booktitle={Medical Imaging 2011: Ultrasonic Imaging, Tomography, and Therapy},
  volume={7968},
  pages={796818},
  year={2011},
  organization={International Society for Optics and Photonics}
}

@article{chernyakova2014fourier,
  title={Fourier-domain beamforming: the path to compressed ultrasound imaging},
  author={Chernyakova, Tanya and Eldar, Yonina C},
  journal={IEEE transactions on ultrasonics, ferroelectrics, and frequency control},
  volume={61},
  number={8},
  pages={1252--1267},
  year={2014},
  publisher={IEEE}
}

@article{mishali2011xampling,
  title={Xampling: Analog to digital at {sub-Nyquist} rates},
  author={Mishali, Moshe and Eldar, Yonina C and Dounaevsky, Oleg and Shoshan, Eli},
  journal={IET circuits, devices \& systems},
  volume={5},
  number={1},
  pages={8--20},
  year={2011},
  publisher={IET}
}

@article{errico2015ultrafast,
  title={Ultrafast ultrasound localization microscopy for deep super-resolution vascular imaging},
  author={Errico, Claudia and Pierre, Juliette and Pezet, Sophie and Desailly, Yann and Lenkei, Zsolt and Couture, Olivier and Tanter, Mickael},
  journal={Nature},
  volume={527},
  number={7579},
  pages={499},
  year={2015},
  publisher={Nature Publishing Group}
}

@inproceedings{Luijten2019beamforming,
  title={Deep Learning for Fast Adaptive Beamforming},
  author={Luijten, Ben and Cohen, Regev and de Bruijn, Frederik J and Schmeitz, Harold AW and Mischi, Massimo and Eldar, Yonina C and van Sloun, Ruud JG},
  booktitle={ICASSP 2019-2019 IEEE International Conference on Acoustics, Speech and Signal Processing (ICASSP)},
  pages={1333--1337},
  year={2019},
  organization={IEEE}
}

@inproceedings{perdios2017deep,
	title={A deep learning approach to ultrasound image recovery},
	author={Perdios, Dimitris and Besson, Adrien and Arditi, Marcel and Thiran, Jean-Philippe},
	booktitle={IEEE International Ultrasonics Symposium (IUS)},
	pages={1--4},
	year={2017},
	organization={Ieee}
}

@inproceedings{gregor2010learning,
  title={Learning fast approximations of sparse coding},
  author={Gregor, Karol and LeCun, Yann},
  booktitle={Proceedings of the 27th International Conference on International Conference on Machine Learning},
  pages={399--406},
  year={2010},
  organization={Omnipress}
}

@article{tur2011innovation,
  title={Innovation rate sampling of pulse streams with application to ultrasound imaging},
  author={Tur, Ronen and Eldar, Yonina C and Friedman, Zvi},
  journal={IEEE Transactions on Signal Processing},
  volume={59},
  number={4},
  pages={1827--1842},
  year={2011},
  publisher={IEEE}
}

@book{eldar2012compressed,
  title={Compressed sensing: theory and applications},
  author={Eldar, Yonina C and Kutyniok, Gitta},
  year={2012},
  publisher={Cambridge University Press}
}

@article{michaeli2012xampling,
  title={Xampling at the rate of innovation},
  author={Michaeli, Tomer and Eldar, Yonina C},
  journal={IEEE Transactions on Signal Processing},
  volume={60},
  number={3},
  pages={1121--1133},
  year={2012},
  publisher={IEEE}
}

@article{mishali2011xampling2,
  title={Xampling: Signal acquisition and processing in union of subspaces},
  author={Mishali, Moshe and Eldar, Yonina C and Elron, Asaf J},
  journal={IEEE Transactions on Signal Processing},
  volume={59},
  number={10},
  pages={4719--4734},
  year={2011},
  publisher={IEEE}
}

@article{gedalyahu2011multichannel,
  title={Multichannel sampling of pulse streams at the rate of innovation},
  author={Gedalyahu, Kfir and Tur, Ronen and Eldar, Yonina C},
  journal={IEEE Transactions on Signal Processing},
  volume={59},
  number={4},
  pages={1491--1504},
  year={2011},
  publisher={IEEE}
}

\end{document}